\documentclass[11pt,fleqn,a4paper,BCOR=18mm,DIV13,cleardoublepage=empty]{article}
\usepackage[english]{babel}
\usepackage{hyperref}
\pdfoutput=1
\usepackage{amsmath}
\usepackage{amssymb}
\usepackage{amsthm}
\usepackage[T1]{fontenc}
\usepackage{graphicx}
\usepackage{latexsym}
\usepackage{ifpdf}
\usepackage{setspace}
\usepackage{color}
\usepackage{caption}
\usepackage{subcaption}
\usepackage{epstopdf}
\usepackage{float}

\newcommand{\dd}{\,\mathrm{d}}

\begin{document}

\title{An Extended Filament Based Lamellipodium Model Produces Various Moving Cell 
Shapes in the Presence of Chemotactic Signals}
\author{Angelika Manhart\footnote{Faculty of Mathematics, University of Vienna,
Oskar-Morgenstern Platz 1,1090 Vienna, Austria,  University of Vienna, Email: angelika.manhart@univie.ac.at, Tel: +431427750716} \and Christian Schmeiser\footnote{ Faculty of Mathematics, University of Vienna, Oskar-Morgenstern Platz 1, 1090 Vienna, Austria, Email: christian.schmeiser@univie.ac.at} 
\and Nikolaos Sfakianakis\footnote{Johannes Gutenberg University,
Staudingerweg 9,
55099, Mainz, Germany,
Email: sfakiana@uni-mainz.de
} \and Dietmar Oelz\footnote{Courant Institute of Mathematical Sciences,
New York University,
251 Mercer Street,
New York, N.Y. 10012-1185,
Email: dietmar@cims.nyu.edu
}}
\maketitle

\begin{abstract}
The Filament Based Lamellipodium Model (FBLM) is a two-phase two-dimensional 
continuum model, describing the dynamcis of two interacting families of locally
parallel actin filaments \cite{Schmeiser2010}. It contains accounts of the
filaments' bending stiffness, of adhesion to the substrate, and of cross-links
connecting the two families.

An extension of the model is
presented with contributions from nucleation of filaments by branching, from capping,
from contraction by actin-myosin interaction, and from a pressure-like repulsion
between parallel filaments due to Coulomb interaction. The effect of a chemoattractant
is described by a simple signal transduction model influencing the polymerization
speed. Simulations with the extended model show its potential for describing
various moving cell shapes, depending on the signal transduction procedure, and
for predicting transients between nonmoving and moving states as well as 
changes of direction.
\end{abstract}


\paragraph{Acknowledgments:}
Financial support by the Austrian Science Fund (FWF) through the
doctoral school \textit{Dissipation and Dispersion in Nonlinear PDEs} (project W1245, A.M.) and the Schr\"odinger Fellowship (J3463-N25, D.O.) as well as the Vienna Science and Technology
Fund (WWTF) (project LS13/029).

\newpage

\section{Introduction}
\label{sec:intro}

In contact to flat adhesive substrates, many cell types tend to develop thin protrusions, called lamellipodia \cite{Small2002}. These are very
dynamic structures, supported by a network of filaments of polymerized actin, which is continuously remodelled by polymerization
and depolymerization as well as by the building and breaking of cross-links and adhesive bonds to the substrate. Polarization of
cells due to internal instabilities \cite{Svitkina1997,Yam2007} or to external signals 
\cite{Postlethwaite1987,Gerisch1981,Iijima2002,Zigmond1973} might lead to crawling movement along the substrate. This movement 
serves as a model for the motility of various cell types in natural environments, such as fibroblasts, tumor cells, leukocytes, keratocytes,
and others. 

The dynamics of the filament network is a complicated process, and effects to be taken into account, additionally to what has been mentioned
above, include the nucleation of new filaments by branching off existing filaments, deactivation of filaments by capping, and contraction by 
actin-myosin interaction (\cite{Lauffenburger1996}, references therein, and more references below). Various attempts have dealt with the formulation of 
mechanical and, consequentially, mathematical models for the 
involved subprocesses as well as for the whole integrated system (\cite{Mogilner2009} and references therein). On the level of individual actin 
filaments, polymerization, depolymerization, 
branching, and capping are typically modeled as stochastic processes, where the regulation of polymerization as the key process pushing the
lamellipodium outward has received the biggest attention \cite{Mogilner1996,Peskin1993}. Models based on individual filaments have provided possible 
explanations for various
phenomena, such as the motility of pathogens in host cells \cite{Mueller2014}. For the description of the morphology dynamics of whole lamellipodia 
these models
are too complex, however. Therefore, continuum models for the mechanics of the filament network have been used, where the choice of model 
is typically guided by the expected rheological properties, such as viscoelasticity or active contraction due to actin-myosin interaction 
\cite{Rubinstein2005,Mogilner2001,Alt1999,Gracheva2004}. 

This work is a continuation of previous efforts \cite{Schmeiser2010} to systematically derive a continuum model from filament based descriptions by an averaging
process similar to homogenization of materials with microstructure. This allows to include detailed knowledge or assumptions on all
subprocesses. We discuss the modeling assumptions of the \emph{Filament Based
Lamellipodium Model} (FBLM), starting with those aspects, which are taken from
\cite{Schmeiser2010} without changes.

\paragraph{Geometry:}
As common in homogenization, the averaging process is facilitated by idealizing assumptions on the microstructure. For 
lamellipodia, the two main assumptions are a restriction to a two-dimensional model motivated by the observed small aspect ratios (100--200 nm
thickness, tens of microns lateral extension, \cite{Small2002}) and the idealization to a network consisting of two families of locally parallel 
filaments crossing each
other transversally. The latter assumption is supported by experimental evidence for steadily moving cells \cite{Winkler2012}. It has to be conceded, 
however, that it 
is questionable in certain conditions such as retracting lamellipodia \cite{Koestler2008}. 

It is assumed that the whole cell is surrounded by the lamellipodium, whose width might vary along the cell periphery. Mathematically speaking, 
the lamellipodium has the topology of a ring lying between two curves, the outer one representing the \emph{leading edge} and the inner one an 
artificially drawn boundary between the lamellipodium and the rest of the cell, roughly defined by a minimum actin density. More precisely,
two non-identical inner boundaries for the two filament families are allowed.

Actin filaments are polar with so called \emph{barbed} and \emph{pointed ends.} All barbed ends are assumed to meet the leading edge 
\cite{Small1978}. 

\paragraph{Filament mechanics:}
Filaments are assumed to resist bending. More precisely, they are modeled as quasi-stationary Euler-Bernoulli beams. They are assumed to be 
inextensible \cite{Gittes1993}.

\paragraph{Cross-links:}
The mechanical stability of the network largely relies on the existence of cross-links between the two families. Candidates for cross-linkers are
proteins such as filamin \cite{Nakamura2007}, but also the Arp2/3 complex providing the connection between filaments at branch points 
\cite{Mullins1998}. It is assumed that cross-linking 
is dynamic with the building and breaking of cross-links as stochastic processes. While attached, cross-links are assumed as elastic, providing 
resistance against relative translational as well as rotational movement (away from an
equilibrium angle) of the two filament families \cite{Schwaiger2004}. Characteristic life times of cross-links are
assumed to be small compared to the dynamics induced by actin polymerization. The corresponding limiting process, which has been carried
out in \cite{Schmeiser2010}, leads to a friction model for the interaction between the filament families.

\paragraph{Adhesion to the substrate:}
Transmembrane protein complexes with integrins as their most important constituent provide adhesive connections between the cytoskeleton 
and the substrate \cite{Li2003,Pierini2000}. Similarly to cross-links it is assumed that these adhesions are transient with relatively small recycling 
times, such that the
averaged effect is friction between the actin network and the substrate. The short life time of adhesions is another questionable assumption,
only satisfied for fast moving cells, where so called \emph{focal adhesions,} i.e. large and very stable adhesion complexes, do not occur.
\bigskip

The FBLM of \cite{Schmeiser2010} is still rather far from a complete description 
of all relevant processes. Some of these gaps are filled by the extensions below.
Most importantly, the total number of filaments is kept fixed and their length
distribution is prescribed in \cite{Schmeiser2010}. Here, filaments will be added by
branching and removed by capping and subsequent decomposition. The length
distribution will be determined by a quasi-equilibrium between polymerization and
severing. In \cite{Schmeiser2010} cell size is regulated by a model for the effect of
membrane tension. Here, this will be replaced by a contractive force in the cell center due to
actin-myosin interaction. In certain applications it might be appropriate to combine
these two effects. A further extension is a little speculative from a modeling point of 
view, but stabilizes the FBLM: We introduce a repulsive effect between parallel
filaments of the same family, motivated by Coulomb interaction caused by the 
significant charges distributed along filaments. Finally, instead of a given fixed
polymerization speed as in \cite{Schmeiser2010}, a model will be formulated involving
both the effect of a chemotactic signal and of local leading edge bending.
More details about these extensions are given in the following paragraphs.

\begin{figure}[H]
\centering
\includegraphics[width=	\textwidth]{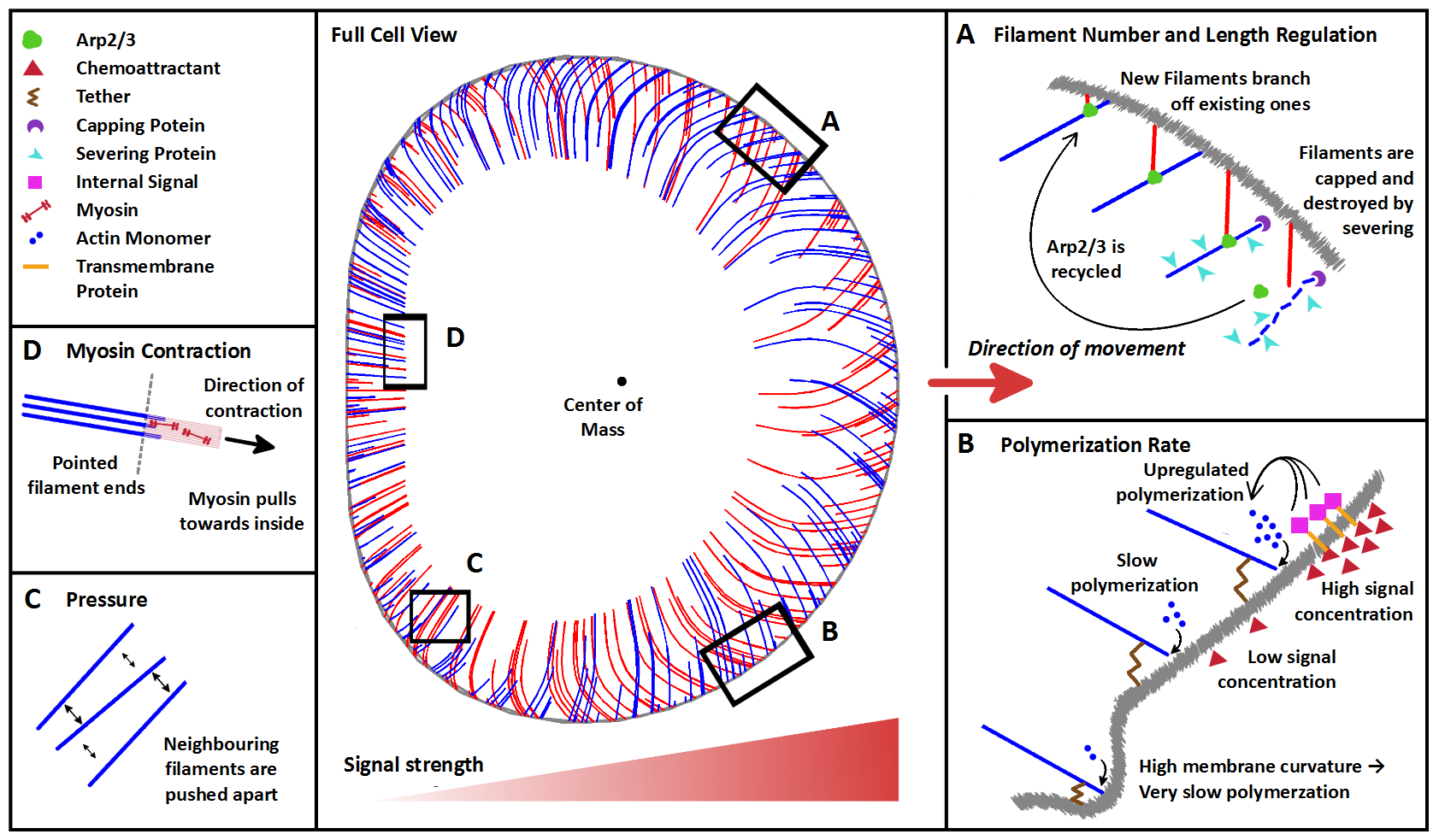}
\caption{\small New ingredients to the FBLM.}
\label{fig:newFBLM}
\end{figure}

\paragraph{Polymerization and degradation:} A desired polymerization speed is
determined between a minimal and a maximal value, depending on the local
concentration of an activator like PIP$_3$, determined by a simple signal transduction
model for a given chemoattractant distribution along the leading edge. The desired
polymerization speed is modified by the pushing force depending on the curvature
of the leading edge (see Fig. \ref{fig:newFBLM}B).

Several degradation processes of filaments are known. 
Aided by proteins of the ADF/cofilin family, they can depolymerize at the pointed ends \cite{Carlier1997}, or bigger pieces of actin can be removed 
by severing proteins like gelsolin \cite{Chaponnier1986}. We assume a severing process (see Fig. \ref{fig:newFBLM}A), a mathematical description of which will lead to formulas for 
the filament length distribution, replacing the ad hoc approximations used in \cite{Schmeiser2010}.

\paragraph{Branching and capping:}
In a lamellipodium, new filaments need to be created in order to maintain a polarized state. New filaments are nucleated by branching off  
existing filaments of the other family at or near to the leading edge ({\em dentritic nucleation model} \cite{Mullins1998,Svitkina1999}). To form a branch 
the presence of the Actin-Related Protein-$2/3$ Complex (Arp$2/3$) at the membrane is necessary. Arp$2/3$ needs to be activated by nucleation promoting 
factors. Activated Arp$2/3$ is incorporated in the branches and later (e.g. upon filament degradation) reenters the cytoplasm, from which it is again 
recruited to the membrane \cite{Machesky1998}. Finally filaments can be capped at their branched ends
by capping proteins \cite{Weeds1993}, which blocks further polymerization (see Fig. \ref{fig:newFBLM}A). The addition 
and removal of filaments had not been taken into account in \cite{Schmeiser2010}.

\paragraph{Confinement:}
As a consequence of polymerization and adhesion, cells would spread indefinitely according to the model components described so far.
In \cite{Schmeiser2010} cell confinement has been modeled as a consequence of membrane tension. However, there is some experimental evidence 
\cite{Svitkina1997} that confinement is mostly due to contractive effects caused by actin-myosin interaction in the rear of the 
lamellipodium. Myosin has the ability to utilize energy to move towards the barbed ends of actin and because of its bipolar structure this leads 
to contraction of antiparallel actin filaments. This mechanism plays an important role in cell movement, because it allows the cell to pull its rear 
parts \cite{Tojkander2012,Mitchison1996,Jay1995,Chen1981}. 

We assume that the artificial inner boundary of the simulated lamellipodium is chosen such that the actin-myosin interaction 
takes place in the interior region not covered by the lamellipodium model. The lamellipodial actin network is assumed to be connected to an
interior network of acto-myosin bundles providing a contractive effect (see Fig. \ref{fig:newFBLM}D). 

\paragraph{Coulomb interaction:}
So far the model assumed no direct interaction between filaments of the same family. Upon trying to understand the bundling of F-actin, it was 
discovered that F-actin, similarly to DNA, behaves like a polyelectrolyte \cite{Tang1997,Tang1996_2,Tang1996}. This means that 
F-actin is negatively charged (about $4 e/nm$) at physiological conditions, hence there exists a repulsive force between the filaments. 
On the other hand, it has been shown that certain positively charged polycations, like divalent metal ions and basic polypeptides 
\cite{Tang1996_2}, which act as counterions and neutralize the negative charges along the filament, promote filament bundling. 

As a modeling assumption, we introduce a repulsive effect between filaments of the same family (see Fig. \ref{fig:newFBLM}C). The consequential 
inhibition of bundle 
formation in the lamellipodium is desired, since this is not our modeling goal at present (although it will be in future work). An additional 
motivation is the fact that a lack of coupling between filaments of the same family may lead to numerical instabilities, which can be avoided 
by the diffusive effect caused by repulsion. \bigskip

The rest of this work is structured as follows. In the following Section \ref{sec:oldmodel} the model of \cite{Schmeiser2010} will be recalled. The new
aspects will be introduced in Section \ref{sec:additions}, and the complete new model will be summarized in Section \ref{sec:newmodel}. 
Finally in Section \ref{sec:results} we will demonstrate the effect of the new terms and the potential of the full new model numerically. The power 
and flexibility of the model will become especially evident in simulations of the polarization 
process induced by a chemotactic signal, of steady movement, and of a turning process. Movies of these simulations including visualizations of the stochastic filament dynamics are contained in the Supplementary Material. Examples of moving cell shapes, influenced by the
response to the chemotactic signal, are shown.

\section{The Filament Based Lamellipodium Model}
\label{sec:oldmodel}

Detailed derivations of the FBLM presented in this section can be found in \cite{Schmeiser2010,Oelz2008}, first simulation results in 
\cite{OelSch-MathBio}, and analytical results for the rotationally symmetric case in \cite{Oelz2010a}. 

In the following, the superscripts $+$ and $-$ refer to the two filament families, also called clockwise and, respectively, anti-clockwise. These superscripts will however be
omitted, whenever we concentrate on one filament family. Quantities related to the
other family will then be indicated by the superscript $*$.

A semi-Lagrangian description of each family is used, where one coordinate $\alpha$ is a filament index varying on a torus 
(represented by $\alpha\in [0,2\pi)$) because of the ring topology, and the other coordinate is the negative arc-length $s$ along a filament 
measured from the leading edge, i.e. $s\in[-L,0]$ with the maximal (simulated) filament length $L$. Because of 
the inextensibility of filaments, $s$ can also be seen as a material coordinate. However, this is only true for fixed time since, by polymerization 
at the leading edge with speed $v(\alpha,t)\ge 0$ and the consequential inward flow of actin relative to the leading edge, a Lagrangian 
coordinate would be $\sigma = s + \int_{t_0}^t v(\alpha,\tau)\dd\tau$. For this reason the material derivative (i.e. the time derivative
at fixed $(\alpha,\sigma)$)
$$
   D_t = \partial_t - v \partial_s
$$
will be used below.

The main unknown in the model is $F(\alpha,s,t)\in\mathbb{R}^2$ which, for fixed $\alpha$ and $t$, represents a
parametrization of the filament with index $\alpha$ at time $t$. By the inextensibility assumption
\begin{equation}
   |\partial_s F(\alpha,s,t)|\equiv 1 \,,\qquad\mbox{for } (\alpha,s)\in B := [0,2\pi)\times[-L,0] \,,\, t\ge 0 \,,\label{eq:inext}
\end{equation}
it is an arc-length parametrization.
More precisely, $F$ represents the deformation of all filaments in 
an infinitesimal $\alpha$-interval, where the length distribution of the filaments is determined by the given function $\eta(\alpha,s,t)$, 
whose value (between 0 and 1) is the fraction of filaments with length at least $|s|$. The assumption that all barbed ends meet the leading 
edge implies that $\eta(\alpha,0,t) = 1$ and that  $\eta$ is monotonically increasing as a function of $s$. Note that 
$\eta(\alpha,s,t)\dd(\alpha,s)$ can be interpreted as the total filament length in the infinitesimal coordinate volume $\dd(\alpha,s)$.

At time $t$, the lamellipodium is represented by the set 
${\cal L}(t) = {\cal L}^+(t) \cup {\cal L}^-(t)$ with ${\cal L}^\pm(t) = \{F^\pm(\alpha,s,t):\, (\alpha,s)\in B\}$.
Note that ${\cal L}^+(t)$ and ${\cal L}^-(t)$ do note have to be identical. We request, however, that they share the leading edge,
which can be motivated by the assumption of tethering of barbed ends to the membrane ("acto-clamping model", \cite{Dickinson2008}): \begin{equation}
 \left\{F^+(\alpha,0,t):0\leq \alpha < 2\pi\right\}=\left\{F^-(\alpha,0,t):0\leq \alpha < 2\pi\right\} \,.\label{eq:barbed}
\end{equation}
The artificial inner boundaries ($s=-L$) might be different.

For the interaction between the two families, the points where filaments cross each other have to be described. This is done on the basis of 
two assumptions:
First, there are no crossings of filaments of the same family, i.e. the map $F^\pm(\cdot,\cdot, t): \,B\to {\cal L}^\pm(t)$ is invertible. 
In particular, we assume that 
\begin{equation} \label{det-sign}
  \det(\partial_\alpha F^\pm,\partial_s F^\pm)>0
\end{equation} 
holds, corresponding to a clockwise parametrization by $\alpha$. 
Second, for each pair of filaments from different families there is at most one crossing, which is transversal.
We need representations of the set ${\cal L}^+(t) \cap {\cal L}^-(t)$ in the coordinate domains. First we identify all pairs of crossing filaments:
\begin{align*}
\mathcal{C}(t):=&\left\{\left(\alpha^+,\alpha^-\right)\in[0,2\pi)^2 : \exists s^\pm(\alpha^+,\alpha^-,t) : \right.\\ 
&\quad \left. F^+(\alpha^+,s^+(\alpha^+,\alpha^-,t),t)=F^-(\alpha^-,s^-(\alpha^+,\alpha^-,t),t)\right\} \,.
\end{align*}
Then we define the parameter domains for both families such that\\ $F^\pm(B_\mathcal{C}^\pm(t),t) = {\cal L}^+(t) \cap {\cal L}^-(t)$:
\begin{align*}
B_\mathcal{C}^\pm(t):=&\left\{\left(\alpha^\pm,s^\pm(\alpha^+,\alpha^-,t)\right):\left(\alpha^+,\alpha^-\right)\in \mathcal{C}(t)\right\}\subseteq B \,.
\end{align*}
Note that, by our assumptions, the transformations
\begin{align}
\label{eq:interact}
 (\alpha^+,\alpha^-)\mapsto \left(\alpha^\pm,s^\pm(\alpha^+,\alpha^-,t)\right)
\end{align}
from $\mathcal{C}(t)$ to $B_\mathcal{C}^\pm(t)$ are invertible and can be combined to transformations
$\psi^\pm:\, B_\mathcal{C}^\mp(t) \to B_\mathcal{C}^\pm(t)$ with the property 
\begin{equation} \label{psi}
  F^\mp = F^\pm\circ\psi^\pm \qquad \mbox{on } B_\mathcal{C}^\mp \mbox{ (for fixed } t).
\end{equation}
The positions and deformations of the filaments are computed on the basis of a quasi-stationary force balance obtained by minimizing a
potential energy functional, which contains contributions from the bending of filaments, the stretching and twisting of cross-links, the stretching
of substrate adhesions, and the membrane tension. This is coupled to age-structured population models for the distributions of cross-links
and adhesions, assuming the building and breaking of these connections as stochastic processes. The resulting model involves continuous
delay terms, since, for the computation of the stretching forces, past deformations of the filaments are needed. With the model in this form,  
numerical simulations would be very expensive, partially also because it mixes different length scales. Whereas the effects of interest occur
on the ($\mu$m) scale of the width of the lamellipdium, the stretching of cross-links and adhesions occurs on molecular (nm) scales.
This implies that motion on the lamellipodium scale of the two filament families relative to each other and relative to the substrate is only 
possible, if the turnover of cross-links and adhesions is fast compared to other mechanisms (e.g. polymerization and depolymerization).

The corresponding limit has been carried out formally \cite{Schmeiser2010} and rigorously for a simplified model problem \cite{Milisic2011}.
It leads to a friction model. The original idea seems to be more than 50 years old and has been formulated first for the derivation of models 
for rubber friction \cite{Schallamach}. Recently it has been used for the modeling of the plastic reorganization of tissues due to cell-cell adhesion
dynamics \cite{PreziosiVitale}. 

The assumption of fast turnover of substrate adhesions is reasonable for fast moving cells such as
fish keratocytes, but certainly not satisfied for focal adhesions, i.e. large stable and long lived adhesion complexes as found, e.g., in fibroblasts.
The limiting procedure removes not only the delay terms, but also the coupling to the population models, which can be solved
explicitly in the limit. 

The variational procedure involves the formulation of a Lagrangian functional where, besides the potential energy, also an account of the
constraints \eqref{eq:inext} and \eqref{eq:barbed} is included. Its variation leads to a weak formulation of the problem for $F$:
\begin{align}
0&=\int_0^{2\pi} \left[\mu^M\left(C-C_0\right)_+\frac{\partial_\alpha F}{|\partial_\alpha F|}\cdot \partial_\alpha (\delta F) 
  \mp \lambda_\text{tether} \nu\cdot \delta F\right]_{s=0}\dd\alpha \nonumber\\
&+\int_{B}\left[\mu^B \partial_s^2F \cdot\partial_s^2 (\delta F) + \mu^A D_t F \cdot\delta F
  + \lambda_\text{inext} \partial_s F \cdot \partial_s (\delta F)\right] \eta \,\dd(\alpha,s) \nonumber\\
& + \int_{\mathcal{C}(t)}\left[\mu^{S}\left(D_t F - D_t^* F^*\right)\cdot \delta F \right. \nonumber\\
 & \qquad\qquad \left. \mp \mu^{T}(\varphi -\varphi_0) \partial_s F^{\perp}\cdot \partial_s(\delta F)\right]
     \eta  \eta^* \dd(\alpha,\alpha^*), \label{weak}
\end{align}
for all variations $\delta F$, where the first line contains contributions from the leading edge, and the convention 
$(F_1,F_2)^\perp = (-F_2,F_1)$ is used. The first term corresponds to
the tension of the membrane with the total length of the leading edge,
\begin{align*}
C:=\int_0^{2\pi}|\partial_\alpha F^+(\alpha,0,t)|\dd\alpha=\int_0^{2\pi}|\partial_\alpha F^-(\alpha,0,t)|\dd\alpha \,,
\end{align*}
its prescribed equilibrium value  $C_0$, and an elasticity coefficient $\mu^M$. The Lagrange multiplier for the constraint \eqref{eq:barbed} 
is a function defined along the leading edge denoted by $\lambda_\text{tether}$, and $\nu$ is the unit outward normal along the leading edge.
The second line of \eqref{weak} deals with forces within individual filaments: resistance against bending with bending modulus 
$\mu^B$, friction with the substrate caused by adhesions with a friction coefficient $\mu^A$, and a tangential force due to the 
inextensibility constraint \eqref{eq:inext} with Lagrange multiplier $\lambda_\text{inext}$. The third and fourth lines of \eqref{weak}
describe the effects of cross-links between the two families with a friction term caused by resistance against stretching of cross-links with
friction coefficient $\mu^S$ and a turning force term caused by resistance of twisting cross-links away from the equilibrium angle 
$\varphi_0$. The angle between the filaments is determined by 
\begin{align*}
\cos\varphi(\alpha^+&,\alpha^-,t) =\\& \partial_s F^+\!\left(\alpha^+,s^+\!\left(\alpha^+,\alpha^-,t\right),t\right)\cdot \partial_s F^-
\!\left(\alpha^-,s^-\!\left(\alpha^+,\alpha^-,t\right),t\right),
\end{align*}
and $\mu^T$ is the corresponding stiffness parameter. The above mentioned limit of fast adhesion and 
cross-link turnover provides explicit formulas for the coefficients $\mu^A$, $\mu^S$, and $\mu^T$ in terms of mechanical and chemical 
properties of adhesion and cross-link molecules.

A strong formulation of the Euler-Lagrange equations requires to transform the domain of the last integral  in \eqref{weak} to $B$.
For this purpose, we use the maps from $\mathcal{C}(t)$ to $B_\mathcal{C}(t)$ described above and introduce the modified friction
coefficient and stiffness parameter
\begin{align}
\label{eq:modstiff}
\widehat{\mu^S}=\left\{
\begin{array}{ll}
\mu^S \left|\frac{\partial{\alpha}^*}{\partial s}\right| & \quad \text{in }  B_\mathcal{C}(t),\\
0 & \quad  \text{else,}
\end{array}\right.\qquad
\widehat{\mu^T}=\left\{
\begin{array}{l l}
\mu^T \left|\frac{\partial{\alpha}^*}{\partial s}\right| & \quad \text{in } B_\mathcal{C}(t),\\
0 & \quad  \text{else.}
\end{array}\right.
\end{align}
The strong formulation is then given by
\begin{eqnarray*}
  0&=&\mu^B \partial_s^2\left(\eta \partial_s^2 F\right) + \mu^A \eta D_t F
    - \partial_s\left(\eta \lambda_\text{inext} \partial_s F\right)\\
&&+ \widehat{\mu^S}\eta\eta^* \Delta V \pm \partial_s\left(\widehat{\mu^T}\eta\eta^* (\varphi-\varphi_0)\partial_s F^{\perp}\right) \,,
\end{eqnarray*}
where the computation of the relative velocity $\Delta V = D_t F - D_t^* F^*\circ\psi^*$ and of the angle between
the families, $\cos\varphi = \partial_s F\cdot (\partial_s F^*\circ\psi^*)$, requires the transformation $\psi^*$ between the 
coordinate domains.

The corresponding boundary conditions are
\begin{align*}
  &-\mu^B\partial_s \left(\eta\partial_s^2F\right) + \eta \lambda_\text{inext} \partial_s F
     \mp \widehat{\mu^T}\eta\eta^* (\varphi-\varphi_0)\partial_s F^{\perp}\\
  &\qquad=\left\{ \begin{array}{ll}  0 & \quad \text{for } s=-L,\\
     \pm\lambda_\text{tether} \nu 
     + \mu^M \left(C - C_0\right)_+\partial_\alpha\left(\frac{\partial_\alpha F}{|\partial_\alpha F|}\right) 
     & \quad  \text{for } s=0,
\end{array}\right.\\
&\eta\, \partial_s^2F =0 \qquad \text{for } s=-L,0.
\end{align*}

\section{Modifications and Extensions}
\label{sec:additions}

\subsection*{Length Distribution and Filament Number Regulation}
\label{ssec:filnr}

In the model of \cite{Schmeiser2010}, the filament number was conserved and the length distribution of filaments was prescribed with a fixed maximum length.
In this section the model will be extended to include the effects of capping, branching, and severing on the filament number and length distribution. 
The results partially depend on the polymerization speed, the choice of which will be discussed below.

 The changes in filament numbers by branching and capping require a different interpretation
of the length distribution $\eta(\alpha,s,t)$. For fixed $s$, $\eta(\alpha,s,t)$ will be considered as the number density of filaments of length at least $-s$
in terms of $\alpha$. Instead of the uniform distribution $\eta(\alpha,0,t)\dd\alpha = \dd\alpha$ of barbed ends, values of
$\eta(\alpha,0,t)$ different from one are allowed. The density of barbed ends per leading edge length is then given by
$$
  \overline\rho(\alpha,t)=\frac{\eta(\alpha,0,t)}{|\partial_{\alpha}F(\alpha,0,t)|}
$$
For the other family, the barbed end density $\overline\rho^*(\alpha^*,t)$ is defined analogously. With $s=0$, the map between the coordinate domains
(see \eqref{psi}) reduces to a map $\alpha^*(\alpha,t)$, and in the following $\overline\rho^*$ means $\overline\rho^*(\alpha^*(\alpha,t),t)$.
In the rest of this subsection we shall deal with fixed values of $\alpha$. The dependence on $\alpha$ will therefore be suppressed for ease of reading.

We start with the evolution of the number of barbed ends and assume that it depends on the barbed end densities per unit length: 
\begin{align}\label{eq:barbed_ends}
&\partial_t \eta(0,t)=f\left(\overline \rho,\overline \rho^*\right)|\partial_\alpha F(0,t)|
\end{align}
where $f\left(\overline \rho,\overline \rho^*\right)$ is the change of barbed end number per unit length and time, modeling the effects of branching and 
capping at the barbed ends. Capped filaments become inactive and are assumed to be depolymerized very fast, such that they can be eliminated from the
system immediately.

It is instructive to rewrite \eqref{eq:barbed_ends} in terms of the length $x = \int |\partial_\alpha F| \dd\alpha$ along the leading edge, instead of the 
Lagrangian variable $\alpha$. With the lateral flow velocity $v_l = \int \partial_t |\partial_\alpha F| \dd\alpha$ (implicitly given as part of the filament dynamics), 
it can be written as
\begin{equation} \label{eq:barbed_ends_Euler}
  \partial_t\overline\rho + \partial_x(v_l \overline\rho) = f \,.
\end{equation}

Branching is assumed to be limited by the availability of activated Arp2/3 complex at the leading edge
with density $a(t)$ (number/leading edge unit length). 
Its equilibrium value in the absence of branching is denoted by $a_0$, the branching rate at equilibrium Arp2/3 density by 
$\kappa_\text{br}$, and the capping rate by $\kappa_\text{cap}$. This leads to the model
\begin{align*}
 f&=\kappa_\text{br}\frac{a}{a_0} \overline \rho^* -\kappa_\text{cap}\overline \rho\,.
\end{align*}
With the rate $c_\text{rec}$ of recruitment and activation of Arp2/3 from the cytoplasm, the evolution of Arp2/3 at the leading edge is 
governed by  
\begin{align*}
  \frac{da}{dt}=c_\text{rec} \left(1-\frac{a}{a_0} \right) - \kappa_\text{br}\frac{a}{a_0} \left(\overline \rho +\overline \rho^*\right) \,,
\end{align*}
The second term reflects the fact that activated Arp$2/3$ is incorporated into branches of both families. 
We assume that the Arp$2/3$ dynamics is fast compared to branching and capping and use the quasi steady (Michaelis-Menten)
approximation
\begin{align*}
  a=\frac{a_0 c_\text{rec}} {c_\text{rec}+\kappa_\text{br}\left(\overline\rho +\overline\rho^*\right)} \,,\qquad 
  f(\overline\rho,\overline\rho^*) = \frac{\kappa_\text{br} c_\text{rec} \overline\rho^*} {c_\text{rec}+\kappa_\text{br}\left(\overline\rho
    +\overline\rho^*\right)} - \kappa_\text{cap} \overline\rho\,.
\end{align*}
The model \eqref{eq:barbed_ends_Euler} has already been used in \cite{Grimm2003} with a prescribed lateral flow velocity and with 
$f=\beta \overline\rho^*/(\overline\rho + \overline\rho^*) - \gamma \overline\rho$.  
A significant difference between our model and \cite{Grimm2003} is in the branching rate: For
$\overline \rho=0$ and small values of $\overline \rho^*$, the branching rate of \cite{Grimm2003} is constant, i.e., not limited by the number of
barbed ends of the other family, while in our model it is approximately $\kappa_\text{br}\overline \rho^*$.

An indication of the qualitative behavior of the model can be obtained from considering a situation, where the barbed end densities
do not vary along the leading edge and are governed by the ODE system
\begin{align}
\label{eq:nucdeg}
\dot{\overline \rho}&=f(\overline\rho,\overline\rho^*) \,,\qquad
\dot{\overline\rho^*}= f(\overline\rho^*,\overline\rho)\,.
\end{align}
It is easily seen that for $\kappa_\text{br}<\kappa_\text{cap}$, i.e. capping exceeds branching, the densities 
converge to 0, otherwise the non-trivial steady state 
\begin{equation}\label{rhobar-equ}
  \overline\rho = \overline\rho^* = \frac{c_\text{rec}}{2}\left( \frac{1}{\kappa_\text{cap}} - \frac{1}{\kappa_\text{br}} \right) \,,
\end{equation}
is stable. In separate work \cite{ManSch_leading-edge} we prove that this qualitative behavior carries over to the corresponding
transport-reaction system with prescribed lateral flow velocities.

The length distribution $\eta(s,t)$ of filaments is influenced by branching and capping through the model \eqref{eq:barbed_ends} for
$\eta(0,t)$, but also directly by capping, which removes whole filaments by our above assumptions. We make the modeling decision
that newly branched filaments are capped preferentially. This means that if branching
exceeds capping, i.e., $f(\overline\rho,\overline\rho^*)\ge 0$, we interpret $f$ as an effective net rate of production of new branches
without further capping, whereas in the opposite case no new filaments are nucleated and $-f$ is an effective rate of capping, affecting
already existing filaments with a probability independent of their length.

Degradation of filaments is also
assumed to be facilitated by the action of severing proteins such as gelsolin \cite{Chaponnier1986} or ADF/cofilin \cite{Roland2008},
cutting filaments at random positions. Similarly to the capping process, we assume that rear parts of filaments, which have been cut off, 
are completely decomposed immediately.

Instead of using $\eta$ as unknown, it is more intuitive to write the model in terms of the density $u(s,t) = \partial_s\eta(s,t)$ of 
filaments with respect to their length $-s$. Following \cite{EdelsteinKeshet1998,Ermentrout1998,Roland2008} for the severing part, 
we use the model
\begin{align*}
  D_t u &= \kappa_\text{sev}\int_{-\infty}^0 \left[ H(s-s')u' - H(s'-s)u\right]\dd s' - \kappa_\text{cap,eff} u \\
  &= \kappa_\text{sev} \left( \int_{-\infty}^s u' \dd s' + su\right)- \kappa_\text{cap,eff} u \,,
\end{align*}
where $u'$ abbreviates $u(s',t)$. Cutting of a filament of length $-s$ at the new length $-s'$ occurs with the rate 
$\kappa_\text{sev}H(s'-s)$ leading to a total severing rate $-\kappa_\text{sev}s$ proportional to the length. This is the simplest possible
model. More elaborate severing rates, e.g., with an account of aging of actin filament subunits \cite{Roland2008}, could be included
easily.

The effective capping rate is given, according to the discussion above, by
$$
  \kappa_\text{cap,eff} = \frac{(-f(\overline\rho,\overline\rho^*))_+}{\overline\rho} \,.
$$
It remains to rewrite the model in terms of $\eta(s,t) = \int_{-\infty}^s u(\sigma,t)\dd\sigma$:
\begin{align*}
&D_t \eta  = \kappa_\text{sev}\,s\,\eta - \kappa_\text{cap,eff}\eta \,.
\end{align*}
For a constant polymerization speed $v$ this equation can be solved explicitly by the method of characteristics:
\begin{align}
\label{eq:density_distribution}
\eta(s,t)=\eta(0,t+s/v)\exp\left(-\frac{\kappa_\text{sev}s^2}{2v}-\int_{t+s/v}^t \kappa_\text{cap,eff}(\tau) \dd\tau\right) \,,
\end{align}
where $\eta(0,t)$ is given as the solution of \eqref{eq:barbed_ends}. Assuming that the changes of the polymerization velocity are
slow compared to the filament dynamics, this formula is a valid approximation also for time dependent velocities $v(t)$.

Finally, the maximum simulation length $L$ of filaments is defined by a cut-off at small actin densities, i.e. a value $\eta_\text{min}$ 
for $\eta$. With the rough approximations of replacing $\eta(0,t+s/v)$ by $\eta(0,t)$ and $\kappa_\text{cap,eff}(\tau)$ by
$\kappa_\text{cap,eff}(t)$, $L$ can be computed explicitly:
\begin{align}
\label{eq:length}
L(t):=-\frac{\kappa_\text{cap,eff}(t)}{\kappa_\text{sev}}+\sqrt{\frac{\kappa_\text{cap,eff}(t)^2}{\kappa_\text{sev}^2}
  +\frac{2v(t)}{\kappa_\text{sev}} \log\left(\frac{\eta(0,t)}{\eta_\text{min}}\right)} \,,
\end{align}
resulting in the time dependent coordinate domain 
\begin{equation}\label{B(t)}
  B(t) = \{(\alpha,s):\, 0\le\alpha < 2\pi,\, -L(\alpha,t)\le s \le 0\} \,.
\end{equation}
Most notably, faster polymerization leads to wider lamellipodia.

\subsection*{Myosin contraction}
\label{ssec:myosin}

We model the effect of interior contractile acto-myosin bundles as forces pulling the (artificial) pointed ends of lamellipodial actin filaments.
Neglecting the effect of substrate adhesion of the interior region, the pulling forces are assumed to add up to zero by the action-reaction
principle.  

The question of an appropriate direction for the pulling forces arises. We consider two scenarios: One where the contractile bundles pull the 
filaments tangentially, and one where they pull towards a central point, chosen as the center of actin mass (without having particularly strong
arguments for this choice). Although one can argue for tangential pulling, which does not perturb the directional order of filaments in the
lamellipodium, this choice has two disadvantages: If filaments get too tangential to the inner boundary of the lamellipodium, tangential pulling 
fails to control the size of the cell. Secondly, tangential pulling is a slightly unstable process, since it might reinforce any small deflections of 
the pointed ends. For those reasons and in order to allow more flexibility of the model, we include a mixture of both choices. 

We again use the notation of dropping the superscript $\pm$ for the family under discussion and using $*$ for the other one. The magnitude of 
the tangential force acting on the filament with index $\alpha$ is denoted by 
$f_\text{tan}(\alpha)$ and that of the centripetal force by $f_\text{in}(\alpha)$. We define $V(\alpha)$ as the normalized vector pointing from the 
center of mass
$$
  C_M := \int_B \eta F\,\dd(s,\alpha) \left( \int_B \eta\,\dd(s,\alpha) \right)^{-1}
$$
to the end $F(\alpha,-L(\alpha))$ of filament $\alpha$. The forces can be 
included in the boundary conditions as
\begin{align*}
 -\mu^B\partial_s\left(\eta\partial_s^2F\right)+\eta \lambda_\text{inext} \partial_s F&
    \mp\eta\, \eta^* \mu^T(\phi-\phi_0)\partial_s F^{\perp} \\
&=\eta \left(f_\text{tan}\partial_s F +f_\text{in} V\right),\qquad s=-L \,.
\end{align*}
We postulate a scalar positive quantity $A$, which measures the size of the contraction effect and which is chosen as $A:=\mu^{IP} (A_c-A_0)_+$
with the area $A_c$ encircled by the lamellipodium and its equilibrium value $A_0$.

The forces $f_\text{tan}(\alpha)$ and $f_\text{in}(\alpha)$ are determined by the conditions that
\begin{enumerate}
\item the total force should be close to the current contractility $A$,
\item it should be split between the tangential and centripetal contributions according to a weight $\gamma\in[0,1]$, and
\item the sum of all forces has to be zero.
\end{enumerate}
Mathematically this is realized by minimizing
$$
  \int_0^{2\pi}\eta(s=-L) \left[\frac{(f_\text{tan} - \gamma A)^2}{\gamma} 
    + \frac{(f_\text{in}-(1-\gamma) A)^2}{1-\gamma}\right] \dd\alpha
$$
with the constraint
\begin{align}
\label{eq:ip_constr}
  \int_0^{2\pi} \eta(s=-L) \left[f_\text{tan} \partial_s F(s=-L) + f_\text{in} V\right] \dd\alpha=0 \,,
\end{align}
giving
\begin{align}
\label{eq:innerpull}
  f_\text{tan}(\alpha)=\gamma A[1\!-\!\mu \cdot \partial_s F(\alpha,-L(\alpha))] \,,\quad f_\text{in}(\alpha)=(1\!-\!\gamma)A[1\!-\!\mu \cdot V(\alpha)] \,,
\end{align}
with
$$
  \mu = \left(\int_0^{2\pi}\!\!\!\! \eta\left[\gamma \partial_s F \otimes \partial_s F +(1-\gamma)V\otimes V\right]\dd\alpha\right)^{-1} \!\!\!\!
    \int_0^{2\pi} \eta\left[ \gamma\partial_s F + (1-\gamma)V \right] \dd\alpha \,.
$$
In Section \ref{sec:results} it will be shown that myosin pulling can effectively control cell size, and that the contraction force allows to produce moving cells. 
For these reasons we neglect the contribution of membrane tension.

\subsection*{Filament repulsion}
\label{ssec:pressure}

We consider a repulsive effect between parallel filaments caused by Coulomb interaction. The presence of mobile charge carriers in the
cytoplasm leads to Debye screening with a typical Debye length in the range of a few nm, such that only local Coulomb interaction can be
assumed, leading to a pressure-like repulsion effect. The electrostatic energy
$$
  U_\text{pressure} = \int_{{\cal L}} \rho\, \Phi\, \dd A
$$
is added to the potential energy, where ${\cal L}(t)$ is the area covered by the filament family under discussion at time $t$ (again dropping the
superscript $\pm$), $\rho\, \dd A$ is the filament length in the infinitesimal area element $\dd A$, with
\begin{align}
\label{eq:rho}
\rho = \frac{\eta}{\det(\partial_\alpha F, \partial_s F)}  \,,
\end{align}
(where (\ref{det-sign}) has been used) and $\Phi$ is the electrostatic potential (see \cite{Ramic2011} for a derivation from a microscopic model based on 
individual filaments). A quasineutral approximation (justified by the relative 
smallness of the Debye length)
and an equilibrium assumption for the mobile charge carriers result in a model for the electrostatic potential of the form $\Phi = \Phi(\rho)$.
As example, the Boltzmann-Poisson model for the mobile carrier density leads to $\Phi = \mu^P \log (\rho)$, $\mu^P > 0$.

For the purpose of computing its variation, the electrostatic energy is written in terms of the quasi-Lagrangian coordinates,
\begin{align*}
&U_\text{pressure}[F] = \int_{B(t)} \Phi(\rho)\eta \,\dd(\alpha,s) \,,
\end{align*}
with the variation in the direction $\delta F$,
\begin{align*}
&\delta U_\text{pressure}[F]\delta F\! =\! -\!\int_B\!\! p(\rho)
\left[\det(\partial_\alpha F,\partial_s (\delta F)) \!+\!
\det(\partial_\alpha (\delta F),\partial_s F)\right]\!\dd(\alpha,s) .
\end{align*} 
The considerations below will show that for stability reasons the pressure $p(\rho) = \Phi'(\rho)\rho^2$ has to be a nondecreasing function of the
density $\rho$, which holds for the Boltzmann-Poisson model $p(\rho) = \mu^P\rho$. Although we expect the pressure only to act in the direction orthogonal to the filaments, this consideration has not entered
the discussion so far. However, the action of the pressure along filaments is eliminated by the (incompressibility) constraint \eqref{eq:inext}.
\smallskip

\noindent{\bf Model problem:} It is instructive to look at a one-dimensional model problem, where points with Lagrangian label $\alpha\in\mathbb{R}$ move along a line with
positions $x(\alpha,t)\in\mathbb{R}$ (assumed a strictly increasing function of $\alpha$). 
The density of points is then given by $\rho = 1/\partial_\alpha x$. The electrostatic energy takes the form $U_\text{pressure}[x] =
\int \Phi(1/\partial_\alpha x) \dd\alpha$. Its ($L^2$-)gradient is given by $\partial_\alpha p(\rho)$. If only the Coulomb interaction
and friction (i.e. adhesions) are taken into account, the dynamics is governed by the gradient flow
\begin{align*}
\partial_t x = - \partial_\alpha p(\rho) \,.
\end{align*}
With the continuity equation (in Eulerian coordinates) $\partial_t \rho + \partial_x(\rho \,\partial_t x) = 0$, this can be rewritten in Eulerian
coordinates as
\begin{align*}
&\partial_t \rho =\partial_x^2 p(\rho) = \partial_x(p^\prime(\rho)\partial_x\rho) \,.
\end{align*}
This is a nonlinear diffusion equation, where nonnegativity of the diffusivity $p^\prime(\rho)$ is necessary for stability. 

For the lamellipodium model, we may hope that the pressure term, by 
causing diffusion in the $\alpha$-direction, avoids intersections within a family and thereby stabilizes the system by ensuring that the modeling assumptions are not destroyed 
by the dynamics. This stabilizing effect is sometimes useful for numerical simulations as demonstrated in Section \ref{sec:results}.

\subsection*{Polymerization Rate}\label{sec:polymerization}

Polymerization rates and, consequentially, polymerization speeds $v(\alpha,t)$ are subject to different regulatory mechanisms. We consider reaction to 
chemotactic signals, where the cell senses concentration gradients of a chemoattractant and translates them to varying polymerization rates. 
This leads to cell polarization and directed movement.
The chemoattractant binds to receptors on the cell membrane that can trigger signaling pathways producing intracellular gradients along the membrane 
reflecting the distribution of occupied receptors. For example, higher concentrations of PIP$_3$ have been observed towards chemotactic signals 
at the leading edge of {\em Dictyostelium discoideum} and of neutrophils \cite{King2009}. This in turn is expected to induce a local upregulation of actin
polymerization \cite{Iijima2002,Weiner1999}. 

We consider constant planar chemoattractant gradients with the chemoattractant concentration $S(x,y) = S_0 + S_1(x\cos(\varphi_\text{ca}) +
y \sin(\varphi_\text{ca}))$, where $S_1$ determines the strength of the gradient and $\varphi_\text{ca}$ its direction. We model a chemotactic response
independent of the strength of the chemoattractant gradient. A normalized internal quantity defined along the leading edge is given by
\begin{align*}
d_\text{ca}(\alpha,t)&=\frac{S(F(\alpha,0,t))-\min\limits_{\beta\in[0,2\pi)} S(F(\beta,0,t))}{\max\limits_{\beta\in[0,2\pi)} S(F(\beta,0,t))
-\min\limits_{\beta\in[0,2\pi)} S(F(\beta,0,t))}
\end{align*} 
Typically, PIP$_3$ is only observed at a part of the leading edge, possibly as consequence of a thresholding phenomenon of the signaling pathway. To account 
for this, we choose a threshold $c\in[0,1]$ and define 
\begin{equation} \label{threshold}
I(\alpha,t)=\left\{
  \begin{array}{l l}
    \frac{d_\text{ca}(\alpha,t)-c}{1-c} & \quad \text{for } d_\text{ca}(\alpha,t)>c \,,\\
    0 & \quad \text{else,}
  \end{array} \right.
\end{equation}
which can be interpreted as a normalized PIP$_3$ concentration. The desired polymerization speed is chosen between prescribed minimal and maximal
values:
\begin{align*}
v_{\text{opt}}(\alpha,t)&=v_\text{min}+I(\alpha,t)\left(v_\text{max}-v_\text{min}\right) \,.
\end{align*}
Finally, the polymerization speed is reduced by the force required to bend the leading edge outwards. On the other hand, due to filament
tethering, we expect some acceleration of polymerization at leading edge segments which are curved inwards. These effects are described
by an ad-hoc model for the polymerization speed $v$, depending on the signed local curvature $\kappa(\alpha)$ (positive for convex leading
edge regions):
\begin{align*}
v=\frac{2v_{opt}}{1+\exp(\kappa/\kappa_{ref})}
\end{align*}

\section{Full New Model}
\label{sec:newmodel}

The functions and variables used are the same as in the original model described in Section \ref{sec:oldmodel}. 
To account for the different 
width of the lamellipodium for different filament indices $\alpha$, we replace $B$ by
\begin{align*}
&B(t):=\left\{\left(\alpha,s\right)| \alpha\in [0,2\pi), s\in[-L(\alpha,t),0]\right\}
\end{align*}
where $L(\alpha,t)$ is given by \eqref{eq:length}, as already introduced in \eqref{B(t)}.\\
\\
The full weak formulation reads 
\begin{align*}
&0=\int_{B(t)}\left(\underbrace{\mu^B \partial_s^2F \cdot \partial_s^2 \delta F}_\text{bending}+\underbrace{\mu^A D_t F
\cdot \delta F}_\text{adhesion} + \underbrace{\lambda_\text{inext}\partial_s F\cdot \partial_s\delta F}_\text{in-extensibility}\right)
\eta \dd(\alpha,s)\\
& - \int_{B(t)}\underbrace{ p(\rho)\Bigl[\det\left(\partial_\alpha F,\partial_s\delta F \right)+\det\left(\partial_\alpha \delta F,\partial_s F\right)\Bigr]}_\text{pressure} \dd(\alpha,s)\\
&+\int_{\mathcal{C}(t)}\!\!\left(\underbrace{\mu^S \left(D_t F-D_t^* F^*\right)\!\cdot\! \delta F}_\text{cross link stretching}
\mp \underbrace{\mu^T \left(\phi-\phi_0\right)\partial_s F^\perp\!\cdot\! \partial_s\delta F}_\text{cross link twisting}\right)\!\eta \eta^* 
\dd(\alpha,\alpha^*)\\
&+\underbrace{\int_{(0,2\pi]} \left(f_\text{tan} \partial_s F+f_\text{in} V \right)\cdot \delta F \eta\Bigm|_{s=-L}
 \dd\alpha}_\text{myosin contraction} 
 \mp \underbrace{\int_{(0,2\pi]}\!\!\!\lambda_\text{tether} \nu \cdot \delta F \Bigm|_{s=0}\!\!\!\dd\alpha}_\text{tethering} \,.
\end{align*}
The filament dependent magnitude of the inner pulling force $f_\text{tan}$ and $f_\text{in}$ are given by \eqref{eq:innerpull} and $\rho$ is defined by \eqref{eq:rho}. The filament densities $\eta(\alpha,s,t)$ are determined by 
\eqref{eq:barbed_ends} and \eqref{eq:density_distribution}.\\

For the strong form, i.e. the Euler Lagrange Equations, the modified stiffness parameters are defined analogously as in \eqref{eq:modstiff}.
\begin{align}
\label{eq:newmodel}
0&=\mu^B \partial_s^2\left(\eta \partial_s^2 F\right)-\partial_s\left(\eta \lambda_\text{inext} \partial_s F\right)+\mu^A \eta D_t F\\
&+ \partial_s\left( p(\rho) \partial_\alpha F^{\perp}\right)
 -\partial_\alpha \left( p(\rho) \partial_s F^{\perp}\right) \nonumber\\
&\pm\partial_s\left(\eta\eta^* \widehat{\mu^T} (\phi-\phi_0)\partial_s F^{\perp}\right)  
 + \eta\eta^* \widehat{\mu^S}\left(D_t F - D_t^* F^*\right) \,,\nonumber
\end{align}
where in the equation for $F^+$, the derivatives of $F^-$ are evaluated at\\ $(\alpha^-(\alpha,s,t), s^-(\alpha,s,t))$ and vice versa. The corresponding 
boundary conditions are
\begin{align}
\label{eq:newBC}
- \mu^B\partial_s&\left(\eta\partial_s^2F\right) - p(\rho)\partial_\alpha F^\perp + \eta \lambda_\text{inext} \partial_s F  
   \mp\eta\eta^* \widehat{\mu^T}(\phi-\phi_0)\partial_s F^\perp \\
&=\left\{
\begin{array}{l l}
\eta \left(f_\text{tan}(\alpha)\partial_s F + f_\text{in}(\alpha) V(\alpha)\right) & \quad \text{for } s=-L \,,\\
 \pm\lambda_\text{tether} \nu & \quad  \text{for }  s=0 \,,
\end{array}\right.\nonumber\\
\eta \partial_s^2&F =0 \qquad \text{for } s=-L,0 \,.\nonumber
\end{align}

\section{Numerical Approach and Simulation Results}
\label{sec:results}

In this section we sketch the numerical method for the solution of \eqref{eq:newmodel}, \eqref{eq:newBC}, described in more detail in 
\cite{MOSS}. Simulations of model problems will demonstrate the effect of the new ingredients to the model introduced in this work.
Finally, the full model is used to simulate the reaction of a cell to a chemotactic signal.

\subsection*{Numerical Method}
\label{ssec:numerics}

The numerical approximation of \eqref{eq:newmodel}, \eqref{eq:newBC} is a formidable task. Filament families as described here are a new
type of continuum, where both analytical and numerical approaches are still in early states of development (see \cite{Oelz2010a,OelSch-MathBio}
for first results). The new numerical method used here will be described only briefly, since it is the subject of parallel work \cite{MOSS}, which we refer to
for details.

Instead of dealing with the time-dependent domains $B^\pm(t)$, the equations for both families are transformed to the rectangular and time-independent
computational domain $[0,2\pi] \times [-1,0]$ by rescaling the variable $s$. A rectangular grid with uniform steplengths is used. The grid lines in the
$s$-direction can be interpreted as computational filaments, each discretized by the same number of grid points, independent of its (time dependent) length.

The strong anisotropy in $\eqref{eq:newmodel}$ is reflected in the choice of the finite element space for the spatial discretization. A tensor product space is 
used, where on each grid cell each component of $F$ is represented by a fourth order polynomial, linear in $\alpha$ and cubic in $s$. These interpolate 
positions and first $s$-derivatives at the nodes. In other words, each computational filament is approximated by a cubic spline with linear interpolation
in between. The finite element space is conforming for the weak formulation of \eqref{eq:newmodel}, being continuous in $\alpha$ and $C^1$ in $s$ and,
thus, a subspace of $H^1_\alpha((0,2\pi), H^2_s(-1,0))$.

An implicit-explicit time discretization is used, based on a linearization.
The evaluation of the interaction terms between the two filament families  
requires approximations of the mappings $\psi^\pm$, derived from \eqref{eq:interact}, in order to represent filaments of one family on the grid of the other. 
The inextensibility constraint has been
implemented by an Augmented Lagrangian approach.

\subsection*{Effects of new model ingredients}
\label{ssec:sim_indiv}

\paragraph{Filament Number Regulation:}
\label{sssec:sim_filnr}
Branching and capping regulate the number of barbed ends. This is of particular importance when the polymerization rate varies along the leading edge.
For an explanation the \emph{lateral flow} phenomenon is needed, i.e. the movement of barbed ends along the leading edge, caused
by polymerization and the inclination of filaments. In our model, the $(+)$-family filaments typically move to the left (relative to the protrusion direction), and
the $(-)$-family filaments to the right. Since bigger polymerization speeds increase the lateral flow, this would lead to filament depletion in regions with higher 
polymerization activity without the regulatory effect of branching and capping.
This typically reduces cell motility (see the paragraph on chemotaxis below and Figure \ref{fig:filnrwhy}).

\begin{figure}[h!]
        \centering
	\includegraphics[width=\textwidth]{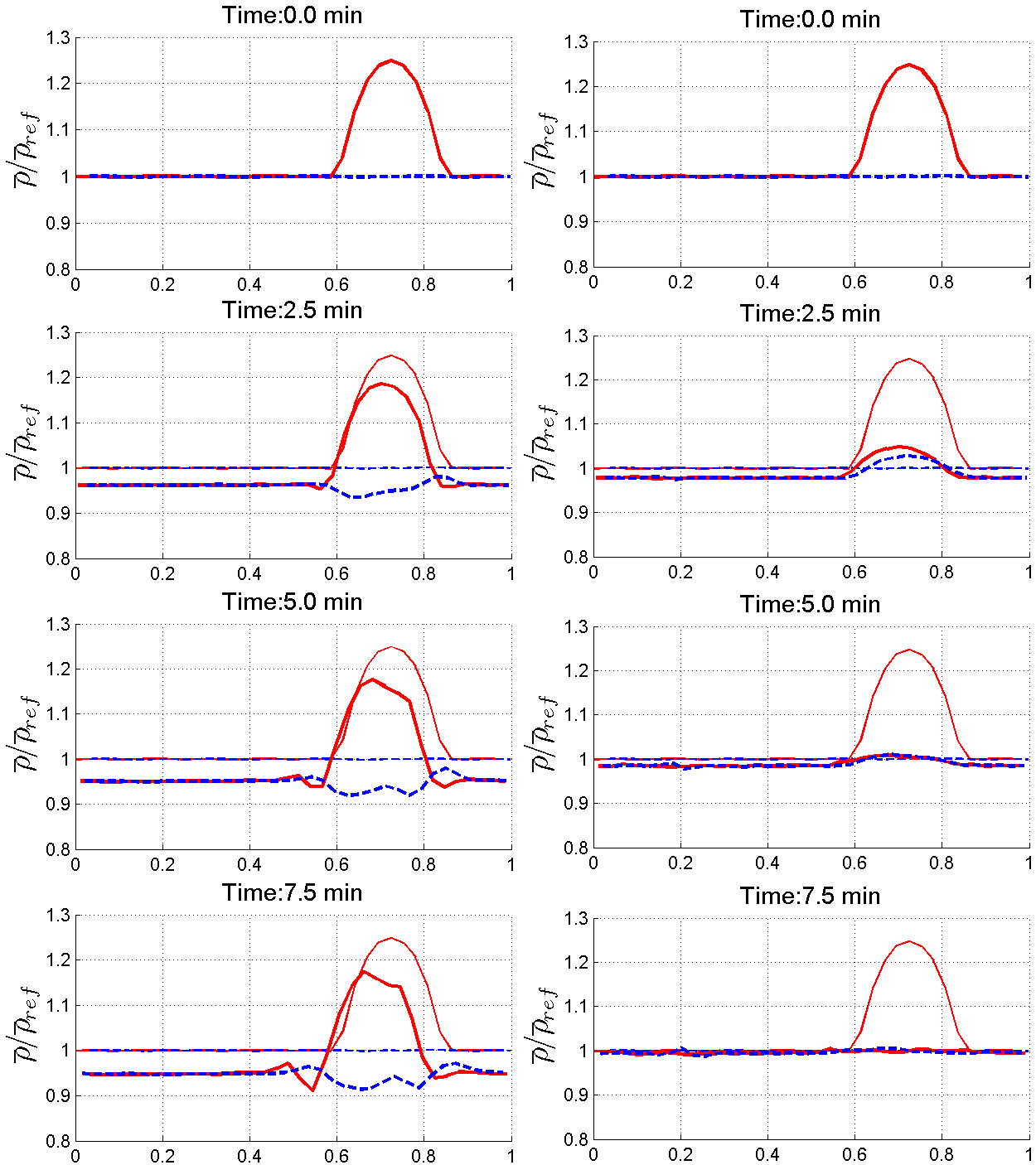}
	\caption{\small Barbed end density perturbation with and without branching/capping: The density of left-moving filaments (red) is perturbed initially, 
	that of right-moving filaments (dashed, blue) not. Thick lines 
	represent the current state, thin lines represent the state at time $t=0$. The left column shows the evolution in the absence of branching 
	and capping, i.e. $\kappa_{br}=\kappa_{cap}=0$. In the right column branching and capping are active with parameters as shown in Table 
	\ref{tab:parameters}, except $\mu^P=0$.}
	\label{fig:filnr}
\end{figure}

To observe the regulation more directly, we started with a radially symmetric cell, where the barbed end density along the leading edge is constant 
and equal to the equilibrium value \eqref{rhobar-equ}. Then the density of barbed ends of the $(+)$-family is perturbed locally. In the subsequent
simulation a constant polymerization speed is used. 
Figure \ref{fig:filnr} shows the evolution of the barbed end densities $\overline \rho^\pm$ for both families.  If there is no branching and 
capping, the perturbation in $\eta^+$ is simply moved to the left by lateral flow without any changes (data not shown). The situation for 
$\overline \rho^+$ is somewhat more complicated, because the geometry changes brought about by the higher number of filaments affects the number of 
filaments per length. Both $\overline \rho^+$ and $\overline \rho^-$ decrease initially because the membrane is locally pushed outward, making the cell 
slightly larger. The lateral flow is also visible here by the shift of the maximum filament number of $\overline \rho^+$.

For the case where branching and capping are active, one can see how initially the number of $(-)$-family filaments increases because of branching. 
Additionally $\overline \rho^\pm$ drops everywhere, again because the cell becomes slightly larger. However the dynamics eventually force the number of 
filaments to return to its equilibrium value everywhere. 

\begin{figure}[h!]
\centering
\includegraphics[width=0.70\textwidth]{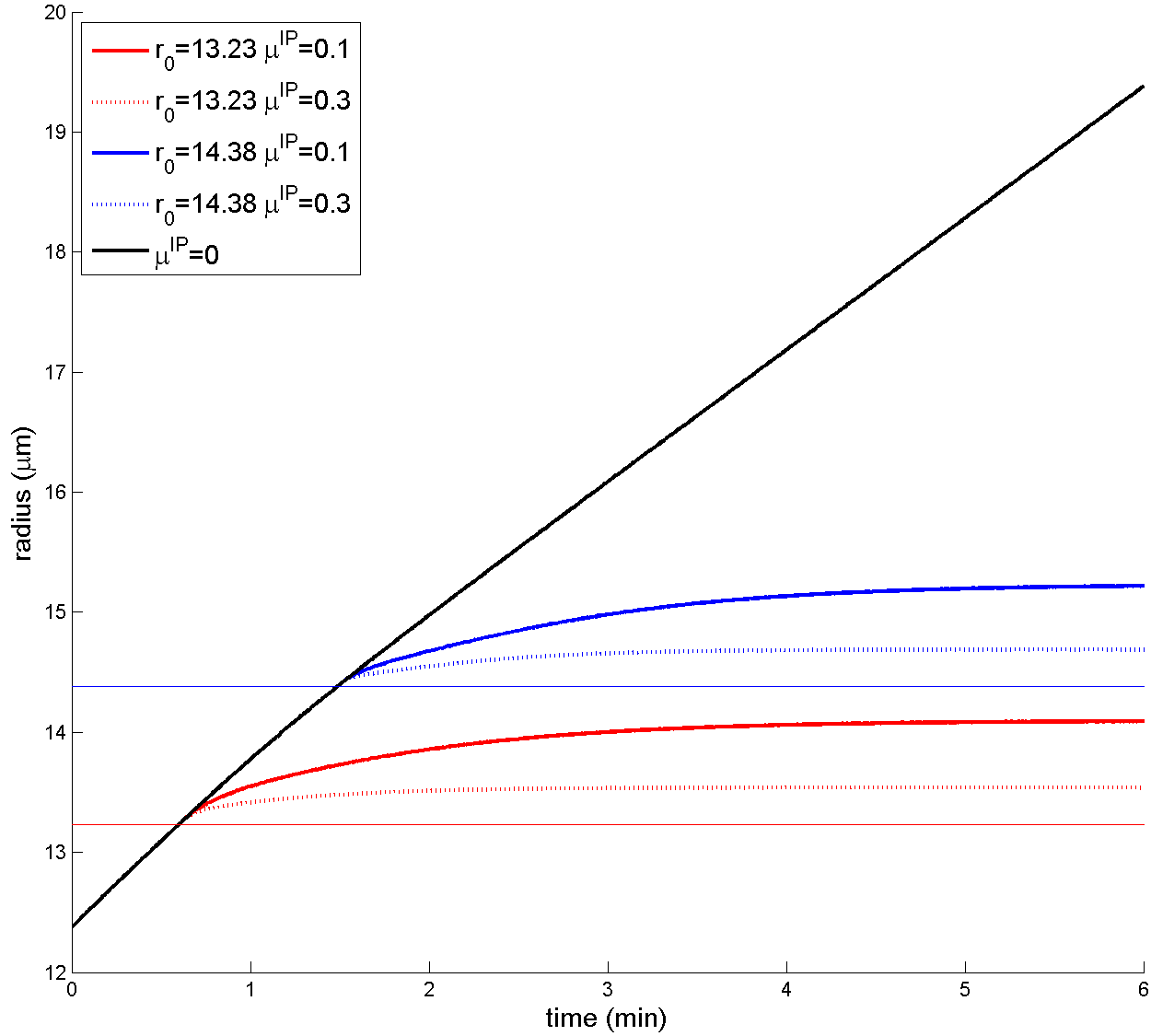}
\caption{\small Effect of the actin-myosin interaction: Comparison of the evolution of the radius of a circular cell for different values of $r_0$ and $\mu^{IP}$. The horizontal lines mark the values of $r_0$. All parameters as in Table \ref{tab:parameters} except the polymerization speed $v_{opt}=0.5$.}
\label{fig:myosin}
\end{figure}

\paragraph{Actin-Myosin Interaction:}
\label{sssec:sim_myosin}
Constraint \eqref{eq:ip_constr} ensures that the myosin pulling on the inside of the lamellipodium is an internal force. We define $A_0=r_0^2\pi$. Figure \ref{fig:myosin} shows the evolution of the inner radius of a 
rotational symmetric cell in time for different values of $r_0$ and $\mu^{IP}$. It can be observed that smaller equilibrium areas and a stronger myosin 
force (i.e. larger $\mu^{IP}$) leads to smaller cell sizes. This shows that the actin-myosin interaction helps to control the cell size. In the section 
on chemotaxis below it will be shown that a moving cell can pull its rear due to this effect.

\paragraph{Pressure:}
\label{sssec:sim_pressure}
The pressure term is a force acting only within one family. To demonstrate its effect we therefore look at a simplified model for one family only:
We assume constant $\eta$, i.e. all filaments of the same length, no polymerization or bending, and only the tangential component of the myosin
force ($\gamma=1$). Then the Euler-Lagrange equations simplify to
\begin{align}
\label{eq:pressure_ex_2}
\mu^A \partial_t F + \partial_s\left( p(\rho)\partial_\alpha F^{\perp}\right) - \partial_\alpha \left( p(\rho) 
\partial_s F^{\perp}\right) - \partial_s\left(\lambda_\text{inext} \partial_s F\right)=0\\
 -p(\rho)\partial_\alpha F^\perp+ \lambda_\text{inext} \partial_s F=\left\{
 \begin{array}{l l}
   0 & \quad \text{for $s=0$}\nonumber\\
    f_\text{tan}(\alpha)\partial_s F & \quad \text{for $s=-L(\alpha)$}
 \end{array} \right.\nonumber
\end{align}
with the constraint $|\partial_s F|\equiv 1$. To enable the computation of an analytical 
solution that can be compared to numerical results, we additionally assume the cell to be rotationally symmetric. This implies that the system can be 
fully described by a reference filament $z(s,t)$. All other filaments are created by rotations. With the rotation matrix $R(\alpha)$ for the rotation angle 
$\alpha$, $F(\alpha,s,t)=R(\alpha)z(s,t)$, leading to $\rho=(z\cdot\partial_s z)^{-1}$.  
Lastly we use the Boltzmann-Poisson model $p(\rho)=\mu^P\rho$. For the pulling force we get $f_\text{tan}=\mu^{IP}(\pi r^2-\pi r_0^2)_+$.

We introduce polar coordinates $z=r \exp(i\, \varphi)$ and denote by $\dot r$ and $r'$ differentiation w.r.t. time and $s$ respectively. This leads to the system
\begin{align}
\label{eq:pressure_ex}
&\mu^A \dot r -\mu^P \left(\frac{r''}{(r')^2} + \frac{1}{r}\right) - \lambda_\text{inext}' r'
  - \lambda_\text{inext}\left(r'' - r(\varphi ')^2 \right) = 0 \,,\\
&\mu^A r\dot \varphi - \lambda_\text{inext}' \varphi' r-\lambda_\text{inext}\left(2 \varphi' r'+\varphi'' r\right)=0 \,,\nonumber\\
&(r')^2+(\varphi')^2 r^2=1 \,,\nonumber\\
& -\frac{\mu^P}{r'} + \lambda_\text{inext} r'=\left\{
  \begin{array}{l l}
    0 & \quad \text{for $s=0$,}\nonumber\\
    \mu^{IP}(\pi r^2-\pi  r_0^2)_+r' & \quad \text{for $s=-L$,}
  \end{array} \right.\nonumber\\
&\lambda_\text{inext} r \varphi'=\left\{
  \begin{array}{l l}
    0 & \quad \text{for $s=0$,}\nonumber\\
    \mu^{IP}(\pi r^2-\pi  r_0^2)_+ r\varphi' & \quad \text{for $s=-L$.}
  \end{array} \right.\nonumber
\end{align}
The first two equations come from the Euler-Lagrange equations, the third is the in-extensibility constraint and the last lines stem from the boundary 
conditions.

\begin{figure}[H]
\centering
\includegraphics[width=	\textwidth]{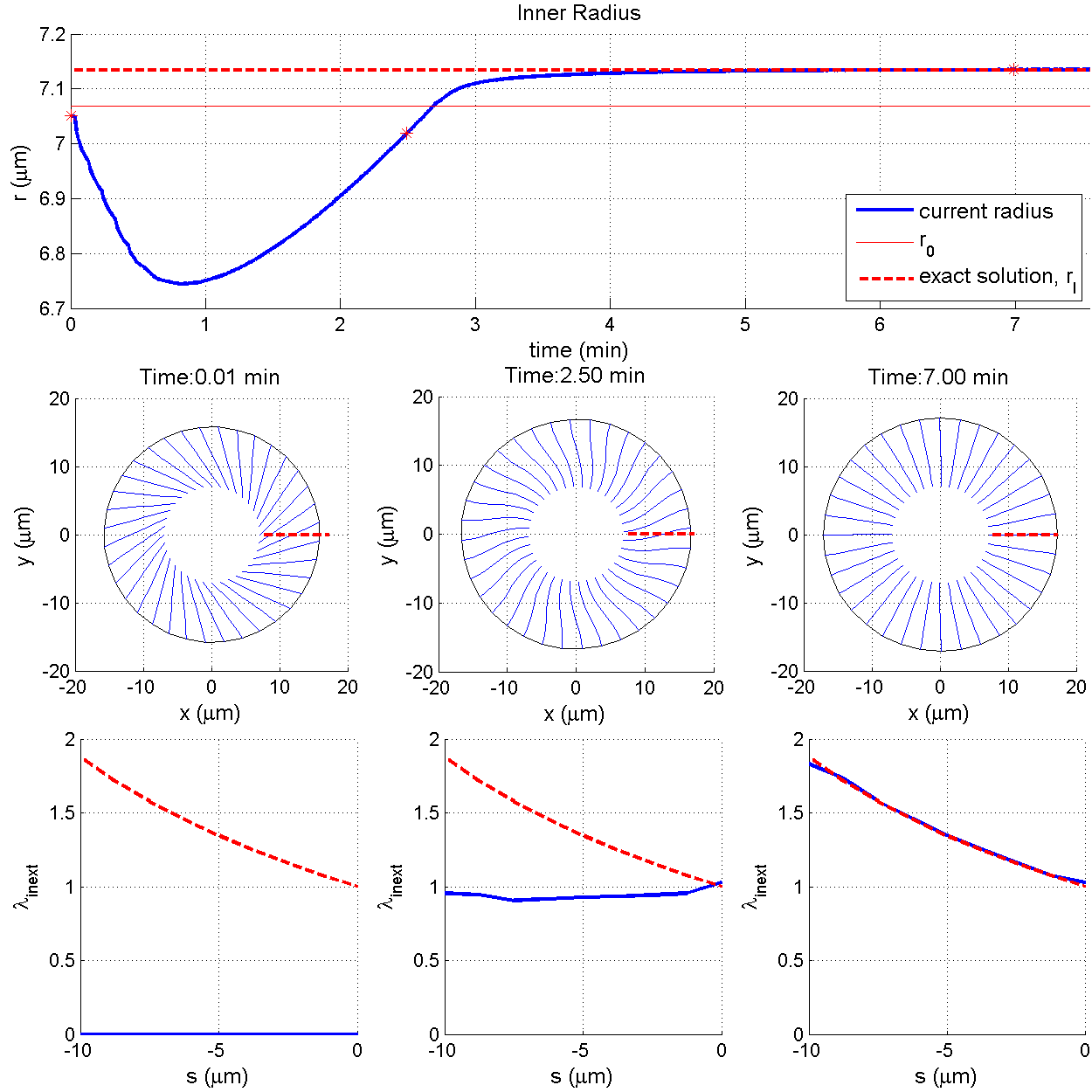}
\caption{\small Pressure. The upper pictures shows the evolution of the inner radius with time. The analytical steady state result $r_I$ (thick, dashed, 
red) and the parameter $r_0$ (thin, red) are also depicted. The stars mark times, where details are shown below: Here the filaments and 
$\lambda_\text{inext}$ at three different times are shown. Again the analytical steady state solutions of the filament positions and of 
$\lambda_\text{inext}$ are shown (thick, dashed, red). $\mu^B=\mu^T=\mu^S=0$, $L=10$, $A_0=r_0^2\pi=157$, $\mu^P=1$. Other parameters as in 
Table \ref{tab:parameters}.}
\label{fig:press}
\end{figure}

A stationary solution with straight radial filaments can be given explicitly:
\begin{align}
\label{eq:pressure_ex_sol}
\varphi(s)&=0\,,\qquad r(s)=s+L+r_I \,,\qquad
\lambda(s)=\mu^P \left( 1 + \ln{\frac{L+r_I}{s+L+r_I}} \right) \,,
\end{align} 
where the inner radius $r_I$ is the unique solution in $(r_0,\infty)$ of
\[
  \ln\left( \frac{L+r_I}{r_I}\right) = \mu^{IP}\pi(r_I^2 - r_0^2) \,.
\]
As initial conditions in a simulation, we take a radially symmetric situation with straight, but not radial 
filaments. We then compare the numerical results to the analytical steady state given by  \eqref{eq:pressure_ex_sol}. Figure \ref{fig:press} shows that convergence
to the steady state is observed. This has been supported by a linearized stability analysis \cite{Leingang}.
Experiment with various initial conditions (data not shown) indicate that there might be an instability with respect to nonsymmetric perturbations. Since
such an instability has never been observed in the full model, it was not investigated further.

\subsection*{Chemotaxis}
\label{ssec:sim_concert}

Finally we want to demonstrate the potential of the full new model, in particular its ability to simulate directed cell movement in the presence of a
chemotactic signal, direction changes, and various cell shapes under different assumptions on the internal signalling network. We want to point out
that we are aiming at a proof of principle and that serious comparisons with and fitting to experimental observations are the subject of ongoing work.
In particular, although our parameter values (discussed below) and results are in realistic ranges, we do not claim the simulated cell shapes
to be close to experiments for any particular cell type.

\begin{figure}[H]
        \centering
	\includegraphics[width=0.75\textwidth]{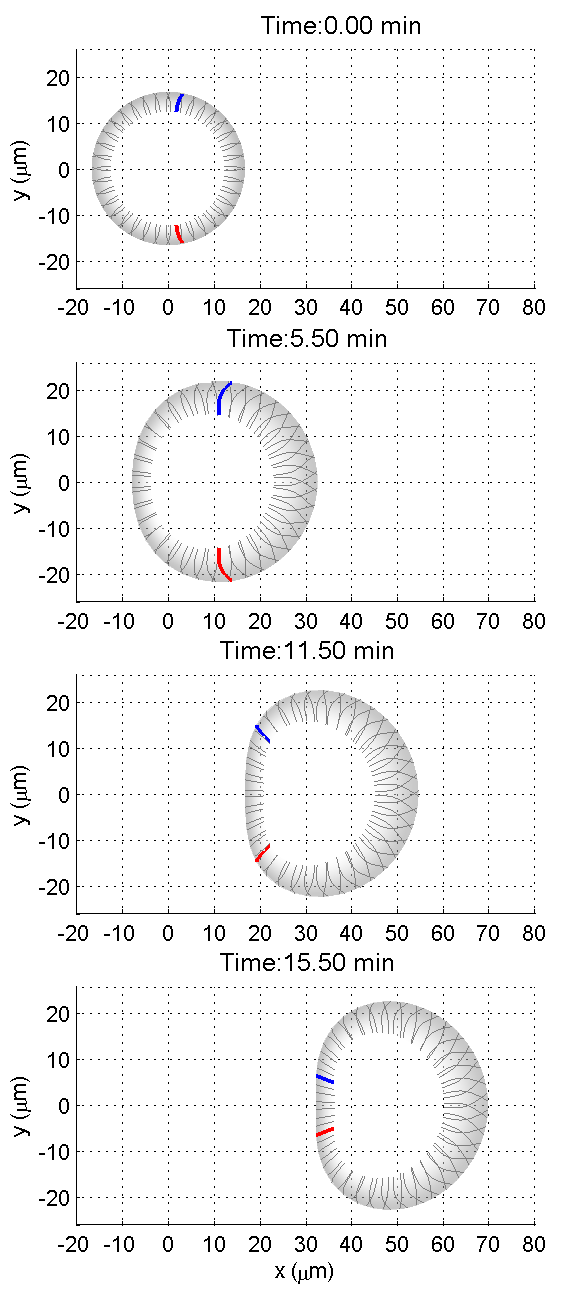}    
\caption{\small Polarization and movement in the presence of a chemical gradient: A time series is shown, where the shading represents 
filament number, thin dashed lines show filament shape, and the thick filaments show the movement of a left moving (red) and right moving (blue) filament 
with time. Parameter values as in Table \ref{tab:parameters} with the internal signal threshold $c=0$. For a movie of the simulation including a visualization of the stochastic filament dynamics see the Supplementary Material.}
	\label{fig:startmov}           
\end{figure}

\begin{figure}[H]    
 	 \includegraphics[width=\textwidth]{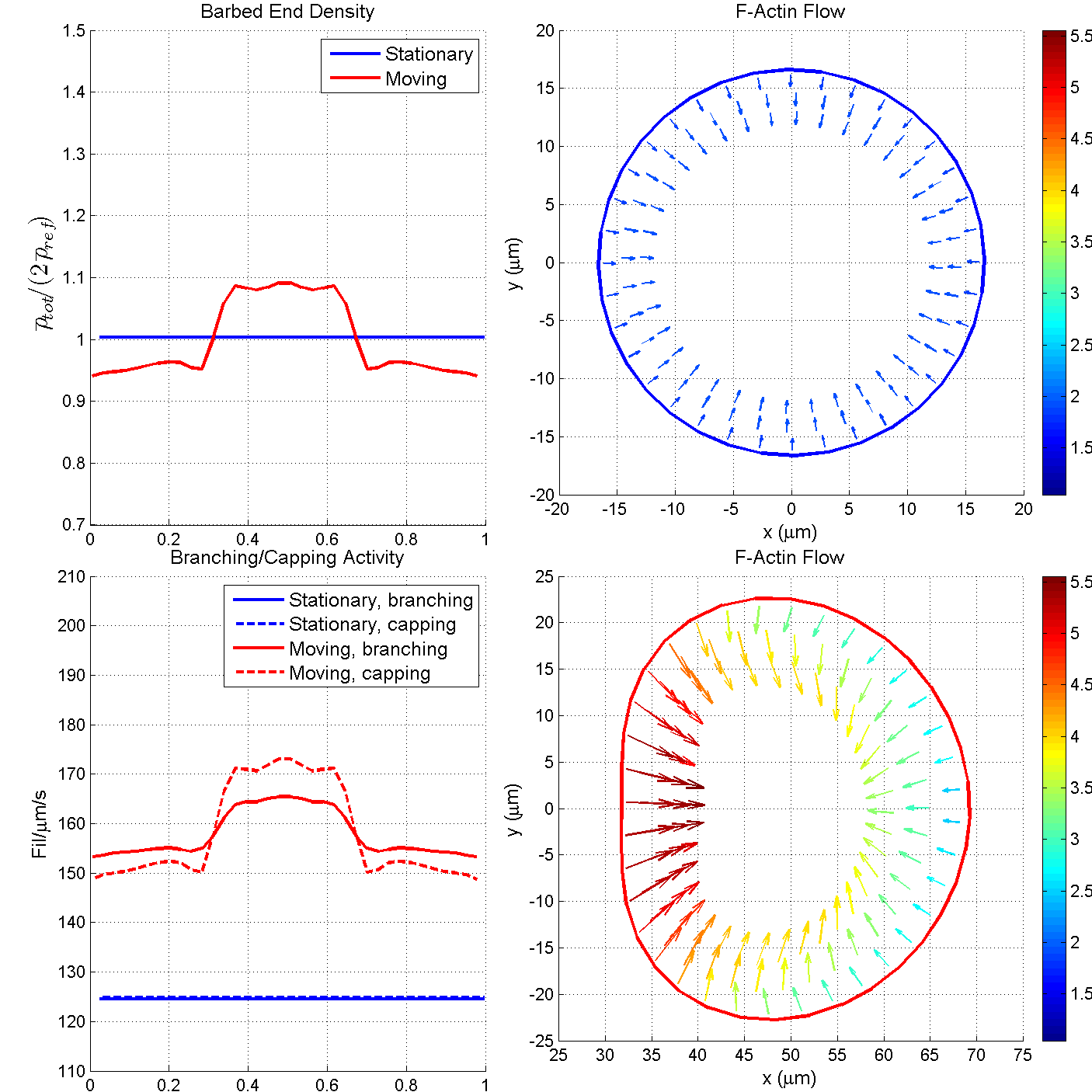}                   
        \caption{\small  Nonmoving vs. moving steady state: Left side pictures show data along the leading edge. On the horizontal axes,
        $0$ and $1$ correspond to the cell front and $0.5$ to the cell rear. The upper 
        left picture shows the barbed end density for the stationary steady state (blue) and the moving steady state (red). The lower 
        left picture shows the branching (solid) and capping (dashed) rates. The two pictures on the right show the F-actin flow field, i.e. the velocity
        of polymerized actin relative to the substrate.  Top: Stationary steady state. Botton: Moving steady state. 
         Arrow length and color (values of the colorbars in $\mu m/min$) represent speed. Parameters as in Table \ref{tab:parameters}.}
	\label{fig:steady}
\end{figure}

\paragraph{Starting to move:}
\label{sssec:sim_startmov}
The following experiment mimics chemotaxis, i.e. a situation where the cell, in reaction to a chemical stimulus (chemoattractant) increases its polymerization rate (\cite{Iijima2002},\cite{Weiner1999}). For the effect of the chemotactic signal on the polymerization speed we use the model introduced in Section
\ref{sec:additions}. A time evolution starting with a rotationally symmetric cell put in a chemical gradient oriented to the right is shown in Figure 
\ref{fig:startmov}. The first visible effect is that the lamellipodium on the right grows wider. The reason is that the increased polymerzation rate 
also increases the maximum filament length as modeled in \eqref{eq:length}. Next the cell starts moving, because filaments grow faster on the right. 
Additionally, the wider lamellipodium at the cell front exerts more friction than the thinner one at the cell rear. Eventually the cell shape 
remains constant and the cell moves steadily towards the right with a speed of about $3.9 \mu m/min$, which is in the biologically observed 
range (\cite{Svitkina1997}, \cite{Iijima2002}). In Figure \ref{fig:steady} the two (numerical) steady states, the stationary cell and the moving cell, 
are shown together with some data. For the stationary cell the number of barbed ends is constant along the leading edge, whereas in the moving cell more 
filaments can be found at the back. To maintain this distribution the cell has to balance branching, capping and movement of filaments. For the 
stationary cell this simply means having branching and capping rates equal everywhere. For the moving cell, branching dominates at the front,
whereas capping exceeds branching in the back. Additionally, the F-actin flow, i.e. the velocity of polymerized actin relative to the substrate is 
depicted in Figure \ref{fig:steady}. For the stationary cell, the flow is rather slow and uniform. 
For the moving cell one can observe small retrograde flow at the cell front and faster flow in the movement direction at the back, where retraction takes place. 
The flow speeds and distributions found are similar to those observed in \cite{Grimm2003}.

\begin{figure}[H]
\centering
\includegraphics[width=1\textwidth]{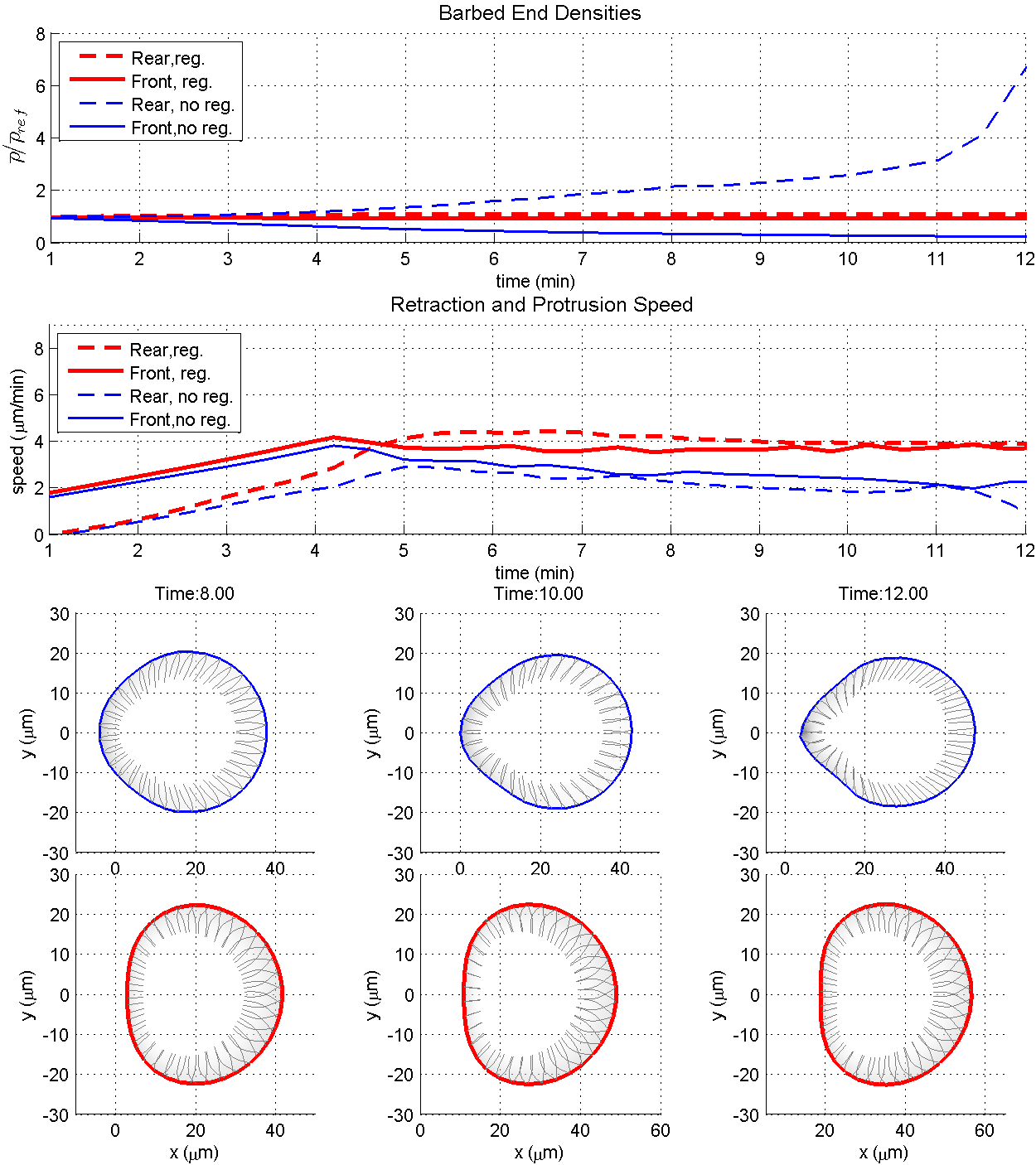}\\
\caption{\small Movement with and without filament number regulation: First row: Retraction speed at the rear (dashed) and protrusion speed (solid) 
at the font of the cell. Red lines (thick) show a cell, which can regulate filament number, blue (thin) lines a cell which cannot. Second row: Same as 
above, but showing the barbed end densities at the rear (dashed) and the front (solid) for the regulated (thick) and unregulated (thin) case. Third and 
fourth row of pictures: Cell shapes at different times. Shading represents actin density, thin gray lines the filament shape. Third row 
(blue, thin leading edge): Cell without regulation. Fourth row (red, thick leading edge): Cell with regulation. Parameters as in Table \ref{tab:parameters}, 
except in the unregulated case $\kappa_{br}=\kappa_{cap}=0$.}
\label{fig:filnrwhy}
\end{figure}

\paragraph{Why filament number regulation?}
\label{sssec:sim_filnrwhy}
Figure \ref{fig:filnrwhy} demonstrates the importance of filament number regulation by branching and capping. The upper picture shows how for a cell with filament number regulation the filament densities at the 
rear and at the front remain close to each other and quite steady over time, whilst in the unregulated case they move away from each other. This is 
because in a cell where regulation is absent, filaments are transported to the back by lateral flow, which leads to an accumulation of filaments at the 
cell rear, whereas the pulling front is depleted of filaments. The middle picture shows how this affects protrusion speed: In the regulated case rear 
retraction and front protrusion speeds approach the same value, as is necessary for constant movement, whilst in the unregulated case rear retraction is 
slower and protrusion velocities decrease with time. In the time series below, one can see that this also affects cell shape: In the unregulated case filaments accumulate in the back, leading to a more prolonged shape.

\begin{figure}[H]
\centering
\includegraphics[width=\textwidth]{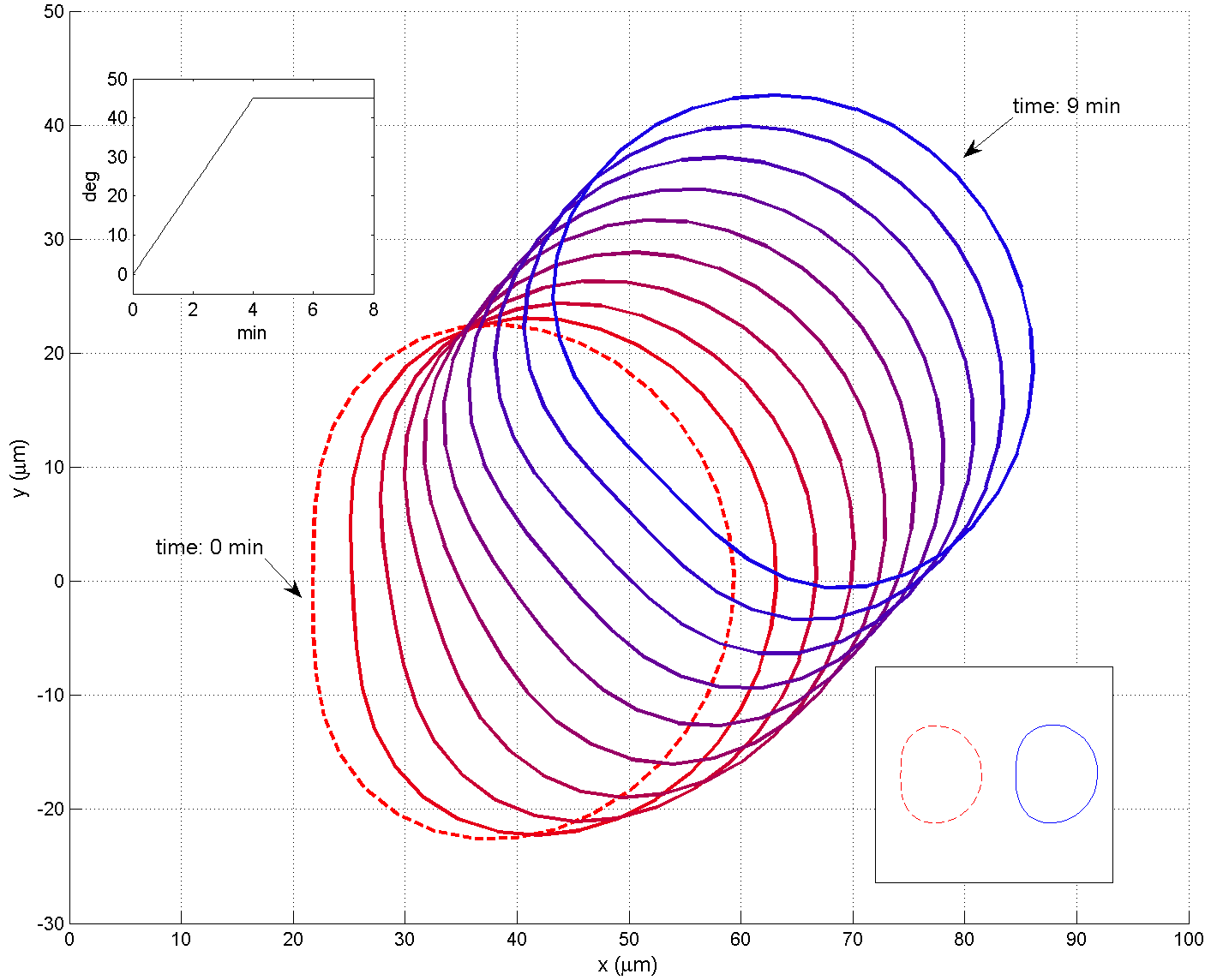}\\
\caption{\small A turning cell. The picture shows the cell shapes of a turning cell over $9$min. Top left inset: The direction of the chemoattractant 
gradient as a function of time. Bottom right inset: The initial state (red, dashed) and the turned final state (blue, solid) are compared.  Parameter values as in 
Table \ref{tab:parameters}. For a movie of the simulation including a visualization of the stochastic filament dynamics see the Supplementary Material.}
\label{fig:turning}
\end{figure}

\paragraph{A turning cell:}
\label{sssec:turning}
One can ask the question if the steady state shape of a moving cell is affected by the initial conditions. A reasonable scenario for investigating this
question, is a situation where the chemoattractant gradient is gradually turned by $45\deg$. Figure 
\ref{fig:turning} shows the corresponding evolution. The final shape is very close to the original shape turned by $45\deg$.

\begin{figure}[H]
        \centering       
                \includegraphics[width=\textwidth]{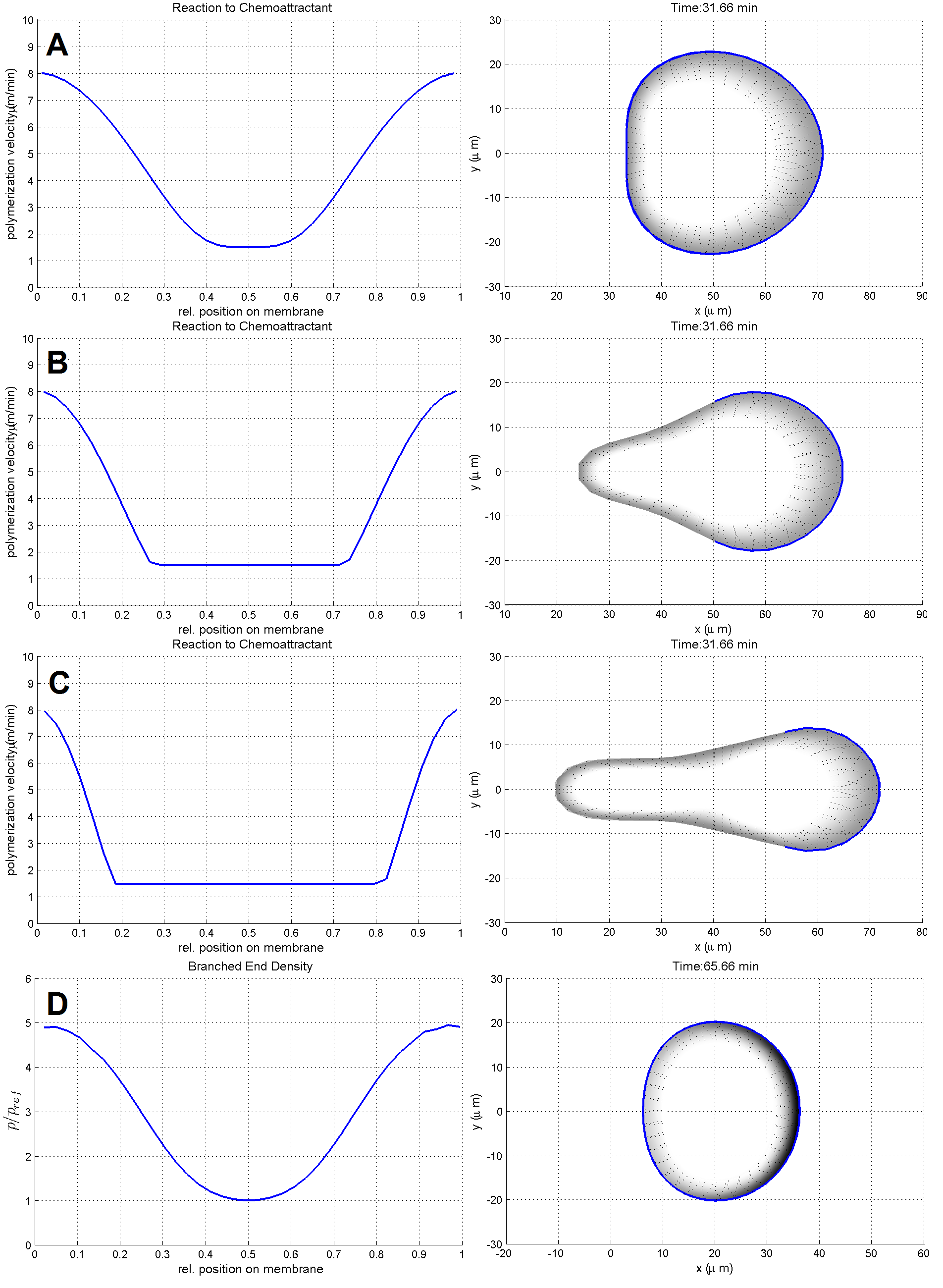}
        \caption{\small Different shapes of moving cells: Left pictures, A--C: Polymerization velocity along the leading edge ($0$ and $1$ representing the cell front and
        $0.5$ the cell rear). Left picture, D: Barbed end density along the leading edge. 
        Right pictures: Final shapes. Shading represents actin density, thin lines in the cells 
        (black, dotted) the filament shapes and lines at the leading edge (thick, blue) indicate the regions affected by the stimulus. A--C: The polymerization rate is affected, A: $c=0$, 
        B: $c=0.5$, C: $c=0.7$. D: The branching rate is affected, $c=0$. Parameter values as in Table \ref{tab:parameters}.}
	\label{fig:shapes}
\end{figure}

\paragraph{Various moving shapes:}
\label{sssec:shapes}
The shape of the moving cell strongly depends on the transduction of the chemotactic signal.
In Figure \ref{fig:shapes}A--C three scenarios are depicted, where only a certain fraction of the leading edge senses the stimulus. The more "local" the effect 
of the stimulus is, the longer the cell gets, because a smaller fraction of filaments pull the cell forward. The differences have been created in the model
by variation of the threshold parameter $c$ in \eqref{threshold}.  The fourth shape in Figure \ref{fig:shapes}D shows results of an alternative mechanism,
where not the polymerization speed but the branching rate is upregulated by the signal. The upregulating mechanism is as in \ref{fig:shapes}A with a
maximally threefold increase of the branching rate leading to a much denser actin network at the front.

\paragraph{Parameters values:}
The parameter values used for the simulation are the ones summarized in Table \ref{tab:parameters} unless stated differently. Where availabe,
values available in the literature have been used (e.g. for $\mu^B, \mu^A, A_0, v^\pm$).  Values for the branching and capping parameters 
$\kappa_\text{br}, \kappa_\text{cap}, c_\text{rec}$ have been chosen in order to give $2\overline \rho_\text{ref}=90$ filaments per $\mu m$ leading edge,
which has been observed in real cells (\cite{Koestler2008}). $\kappa_\text{sev}$ has been chosen in order to give a lamellipodial width in the relevant range of several $\mu m$ \cite{Small2002}. 
Other parameters however ($\mu^S, \mu^T, \mu^P, \mu^{IP}$, $\kappa_{ref}$) result from averaging processes in the derivation of the model
(see \cite{Schmeiser2010}). In principle they can be derived from molecular properties, but they depend in a complicated way on quantities, where not much experimental data
is available, such as mechanical properties of cross-linker molecules, their binding and unbinding rates, and their densities. These parameters can be used
in a fitting process. Values within reasonable ranges have been chosen here.

\begin{table}[H]
\caption{Parameter Values}
\label{tab:parameters}       

\begin{tabular}{| c | p{4cm} | p{4cm} | p{4cm} | }
\hline
Var. & Meaning & Value & Comment \\ 

\hline
& & & \\
$\mu^B$ & bending elasticity & $0.07pN\mu m^2$ &\cite{Gittes1993}  \\
$\mu^A$ & macroscopic friction caused by adhesions & $0.14 pN min \mu m^{-2}$ & measurements in \cite{Li2003,Oberhauser2002}, estimation and calculations in \cite{Oelz2008,Oelz2010a,Schmeiser2010}\\
$\kappa_{br}$ & branching rate & $10 min^{-1} $ & order of magnitude from \cite{Grimm2003}, chosen to fit $2 \overline \rho_\text{ref}=90 \mu m^{-1}$ \cite{Small2002}\\
$\kappa_{cap}$ & capping rate & $5 min^{-1} $ & order of magnitude from \cite{Grimm2003}, chosen to fit $2 \overline \rho_\text{ref}=90 \mu m^{-1}$ \cite{Small2002}\\
$c_{rec}$ & Arp$2/3$ recruitment  & $900\, \mu m^{-1} min^{-1} $ & chosen to fit $2 \overline \rho_\text{ref}=90 \mu m^{-1}$ \cite{Small2002}\\
$\kappa_{sev}$ & severing rate & $0.38 min^{-1} \mu m^{-1}$ & chosen to give lamellipodium widths similar as described in \cite{Small2002} \\
$\mu^{IP}$ & actin-myosin interaction strength& $0.1 pN \mu m^{-2}$ &  \\
$A_0$ & equilibrium inner area & $450 \mu m^2$ & order of magnitude as in \cite{Verkhovsky1999,Small1978}\\
$v_\text{min}$ & minimal polymerization speed & $1.5\mu m/min^{-1}$ & in biological range\\
$v_\text{max}$ & maximal polymerization speed & $8\mu m/min^{-1}$ & in biological range\\
$\mu^P $ &pressure constant & $0.05 pN \mu m $ & \\
$\mu^S$ & cross-link stretching constant & $7.1\!\times\! 10^{-3} pN\, min\, \mu m^{-1} $& \\
$\mu^T$ & cross-link twisting constant & $7.1\times 10^{-3} \mu m$ &\\
$\kappa_{ref}$ & reference leading edge curvature for polymerization speed reduction & $(5\,\mu m)^{-1}$ & \\
\hline
\end{tabular}
\end{table}

\bibliographystyle{plain}
\bibliography{modpaper_bibliography}   

\end{document}